# Wave-induced biases in ADCP measurements from quasi-Lagrangian platforms

A. Y. Shcherbina, and E. A. D'Asaro

*Applied Physics Laboratory, University of Washington, Seattle, WA*

Corresponding author: Andrey Shcherbina, shcher@uw.edu

ABSTRACT

Compact autonomous marine vehicles, both surface and submersible, are now commonly used to conduct observations of ocean velocities using Acoustic Doppler Current Profilers (ADCPs). However, in the inevitable presence of surface waves, ADCP measurements conducted by these platforms are susceptible to biases stemming from wave-coherent orbital motion and platform tilting. In typical ocean conditions, the magnitude of the bias can reach tens of centimeters per second. This paper presents analytical derivation of the depth-dependent bias formulas in the small-amplitude linear wave approximation. A variety of scenarios are considered, encompassing surface and subsurface platforms, upward- and downward-looking ADCPs, free-drifting and self-propelled vehicles. The bias is shown to be a function of the wave field properties, platform response dynamics, and the ADCP configuration (particularly, orientation and beam angle). In all cases, the wave-induced biases show parametric scaling similar to that of the Stokes drift, albeit with a number of critical nuances. Analytical derivations are validated with a semi-analytical model, which can also be used to estimate the biases for more complex measurement configurations and fully nonlinear waves. Further analysis reveals unexpected fundamental differences between the upward- and downward-looking ADCP configurations, offering insights for experimental design aimed at minimizing and mitigating wave-induced biases in autonomous oceanographic observations.

## 1. Introduction

Small autonomous and remotely controlled marine vehicles are being increasingly used as platforms for ocean velocity measurements, typically conducted with acoustic Doppler current profilers (ADCPs). A few examples of such platforms include self-propelled Autonomous Underwater Vehicles (AUVs, e.g., Fong and Jones 2006; Amador et al. 2015); buoyancy-driven underwater gliders (Ellis et al. 2015; Todd et al. 2017); various Autonomous Surface Vehicles (ASVs) such as Waveglider (Mullison et al. 2011; Grare et al. 2021), Saildrone (Gentemann et al. 2020), and Jetyak (Kimball et al. 2014); free-drifting (Lagrangian) underwater floats (Shcherbina et al. 2018) and surface buoys (Thomson et al. 2015); moored profilers (Rusello et al. 2011); and various towed bodies (Münchow et al. 1995; Klymak and Gregg 2001; Hughes et al. 2020), among many others. All these platforms have one critical aspect in common: Due to their relatively small size, their motion is significantly affected by the orbital velocities of the surface waves. To highlight this aspect, we will refer to these platforms, collectively, as "quasi-Lagrangian." This designation would also encompass the idealized fully Lagrangian platforms that follow the wave motions exactly.

While quasi-Lagrangian platforms offer numerous advantages, their wave-coherent motion introduces a particular class of biases into the ADCP measurements. Inherently, ADCPs are remote-sensing instruments, with the sampling volume (for each cell and each beam) some distance from the instrument itself. This sampling volume moves through the water due to the wave-induced coherent motion and tilting of the platform. As a result of this movement, sampling of the wave orbital velocities is conducted in a phase-coherent manner that makes the wave contribution aperiodic. Therefore, the measurement averages deviate from the Eulerian means by what can be called "wave-induced bias." As discussed below, the genesis of this bias is inherently related to that of the Stokes drift, so their physical parameterizations are similar as well.

The geometry of ADCP velocity measurements causes additional complications in the presence of waves. Most ADCPs derive 3D velocity vectors from 3−5 beams under the assumption of horizontal homogeneity of the measured velocity fields. This assumption is easily violated by the wave orbital motions that commonly have horizontal scales close to the spread of the ADCP beams. As a result, measurements of both the wave orbital velocities and the wave-induced biases are affected.

The problem of wave-induced biases has been studied in the past. It was first recognized by Pollard (1973), who considered moored velocimeter biases induced by the motion of the surface float. Amador et al. (2015, 2017) conducted a comprehensive study of wave-induced biases affecting self-propelled autonomous underwater vehicle ADCP measurements using a semi-analytical simulation, observations, and theory. Thomson et al. (2019) focused on a surface Lagrangian platform (SWIFT) and extended the expression of the wave-induced bias to the broadband wave forcing case. On a separate investigation path, Gargett et al. (2008, 2009) derived analytical expressions for the ADCP response to partially-coherent "turbulent" flows arising from the beam separation. These seminal studies were more general in some aspects (e.g., they considered arbitrary phase relationship between the horizontal and vertical velocity components and various ADCP beam patterns), but less general in others (only stationary upward-looking ADCPs were considered).

Here, we bring together these research strands and expand them to consider three distinct mechanisms that lead to wave-induced biases in velocity measurements from quasi-Lagrangian platforms:

1) Translational motion bias: This bias results from the wave-induced orbital motion of the sampling volume.
2) Rotational motion bias: This bias arises from the wave-induced tilting of the platform, which causes a "sweeping" motion of the sampling volume.
3) Frame rotation bias: This bias may occur if the velocity measurement and averaging are conducted in an instrument coordinate frame that is tilted by wave action.

To our knowledge, the latter two biases have not been previously discussed in the literature. We also explore how these biases are affected by the ADCP beam pattern and platform propulsion (horizontal or vertical).

In some cases, we re-derive existing formulations under a common framework. In others, we investigate new aspects of this problem, such as the platform tilt effects (Section 3.3) or the marked asymmetry between the upward- and downward-looking ADCPs (Section 4). And in one case, involving the self-propelled platform motion (Section 6), we argue that previous studies are not quite correct.

In our analysis, we combine analytical derivation with numerical semi-analytical models. As discussed below, each approach has its advantages, and together they serve as an important safeguard against miscalculations.

The paper is organized as follows: Section 2 describes the analytical framework and the semi-analytical model used in our analysis. Section 3 derives expressions for the biases arising from the motion and tilt of a platform affected by a monochromatic deep-water wave. Section 4 discusses the effects of the ADCP beam geometry and how they affect the wave-induced biases. Section 5 considers the wave response of self-propelled platforms and derives necessary corrections to the wave-induced bias formulas. Section 6 briefly discusses the wave-induced bias formulations for the finite-depth wave equations. Section 7 considers the effects of broadband wave forcing. Section 8 discusses how the biases can be mitigated. Section 9 concludes with a brief summary.

## 2. Methods

### *2.1. Analytical framework*

We consider a quasi-Lagrangian platform conducting velocity measurements in the presence of surface gravity waves. For simplicity, we postulate a monochromatic deep-water linear wave with the amplitude $a$, wavenumber $k$, and cyclic frequency $\omega$ propagating in the $x$ direction. Its surface elevation is given by

$$\eta = a \sin \phi, \tag{1}$$

where $\phi = kx - \omega t$ is the wave phase. The wave amplitude $a$ is considered to be small compared to the wavelength, so that the wave steepness parameter $(ak) \ll 1$. For linear deep-water waves, the dispersion relationship requires that $\omega^2 = gk$, where $g$ is gravitational acceleration (finite-depth extension will be considered in section 8). The phase speed of the wave is $c_{ph} = \omega k^{-1}$.

Orbital motions of fluid particles in $x - z$ plane are described as

$$\begin{aligned} x &= x_0 + x_0' = x_0 + a\, e^{kz_0} \cos \phi_0 \,, \\ z &= z_0 + z_0' = z_0 + \, a\, e^{kz_0} \sin \phi_0 \,, \end{aligned} \tag{2}$$

where $\phi_0 = kx_0 - \omega t$ is the wave phase at the mean particle position $(x_0, z_0)$.

For convenience, we express coordinates in the $x - z$ plane using complex notation, $X = x + iz$. With this notation, equations for the particle trajectory (2) can be written as

$$X = X_0 + X_0' = X_0 + a\, e^{i(kx_0 - \omega t) + kz_0} = X_0 + a\, e^{i\phi_0 + kz_0}. \tag{3}$$

This expression highlights the utility of complex notation: As written, the wave orbital excursion $X_0'$ is exponential in all its arguments $(x, z, t)$, therefore it is proportional to all its temporal and spatial derivatives (perhaps with a complex factor). As will become evident shortly, this leads to a significant simplification of the algebraic derivations.

The corresponding wave velocity field, in complex notation, $U = u + iw$, is given by

$$U_0 \equiv U(x_0, z_0, t) = \frac{\partial X}{\partial t} = -i\omega X_0' = -ia\omega\, e^{i(kx_0 - \omega t) + kz_0} = -ia\omega\, e^{i\phi_0 + kz_0}. \tag{4}$$

Or, in components,

$$\begin{aligned} u_0 &= \ \ a\omega e^{kz_0} \sin \phi_0 \,, \\ w_0 &= -a\omega e^{kz_0} \cos \phi_0 \,. \end{aligned} \tag{5}$$

In our calculations we use linear Taylor-series expansion of the velocity field in the vicinity of some point $(X_0 = x_0 + iz_0)$:

$$U(X_0 + \Xi', t) \approx U_0 + x' \left(\frac{\partial U_0}{\partial x_0}\right) + z' \left(\frac{\partial U_0}{\partial z_0}\right), \quad (6)$$

where $\Xi' = x' + iz'$ is an arbitrary small ($|\Xi'| \lesssim a \ll k^{-1}$) deviation from $X_0$. Since

$$\begin{aligned} \frac{\partial U_0}{\partial x_0} &= ikU_0 = \omega k X_0', \\ \frac{\partial U_0}{\partial z_0} &= kU_0 = -i\omega k X_0', \end{aligned} \quad (7)$$

the linear expansion simplifies to

$$\begin{aligned} U(X_0 + \Xi', t) &\approx U_0 + x'(\omega k X_0') + z'(-i\omega k X_0') = \\ &= U_0 + \omega k X_0'(x' - iz') = \\ &= U_0 + \omega k X_0' \Xi'^*, \end{aligned} \quad (8)$$

where an asterisk represents complex conjugate, $\Xi'^* = x' - iz'$. This expansion corresponds to the first-order approximation in the small parameter $(|\Xi'|k)$, which is of the same order of magnitude as the wave steepness parameter $(ak) \ll 1$.

### *2.2. Semi-analytical model*

To validate the analytical calculations and investigate more complex cases, we developed a semi-analytical model of a quasi-Lagrangian velocity-measuring platform. The model assumes the same velocity field associated with surface waves (4) − (5), extended to account for multiple spectral components $a(\omega, k)$, if required. Platform motion is simulated according to (3), and the platform tilt is modeled based on the wave properties and a specified platform response function (see section 3.3). The velocity field is then sampled with multiple “beams” and “cells,” and processed as it would be with an ADCP. This model is semi-analytical, in that the trajectory calculation and velocity sampling is analytical, but the “post-processing” is numerical.

The model does not rely on linear expansion of the velocity field (6) and can therefore handle greater excursions of the sampling volume. Similarly, the model does not rely on small-angle approximation of platform tilt and can therefore handle arbitrary wave-induced variations in platform orientation. Furthermore, the semi-analytical model can be easily extended to incorporate finite-depth linear or non-linear approximations of surface and internal wave motions.

Using the model, various scenarios of platform motion can be also explored. For example, Amador et al. (2015) used a similar model to investigate wave-induced biases in AUV-based velocity measurements. Different algorithms of velocity post-processing, averaging, and referencing can be evaluated as well. The model is implemented in MATLAB and available freely at https://github.com/shcher2018/wave-bias.

## 3. Wave-induced biases

### *3.1. Origins of the bias*

Wave-induced biases considered in this section share a common origin: the biased sampling of the wave orbital velocities by a moving sampling volume. This phenomenon can be illustrated by considering the sampling volume trajectories in the phase space of the wave $(\phi, z)$, where $\phi = kx - \omega t$. As follows from (4), wave orbital velocities are periodic in $\phi$ and grow exponentially with $z$ (Fig. 1b). The velocity anomalies relative to those at a nominal sampling level $z = z_1$ form an asymmetric quadrupole pattern (Fig. 1c).

Sampling volume trajectories will be discussed in greater detail below, but a few generalities can be mentioned upfront. In physical space (Fig. 1a), wave-induced motions of the sampling volumes typically trace circular or elliptical paths around the nominal sampling location, although more complex shapes are possible. In the wave phase space (Fig. 1b, c), the trajectories are oscillatory, resembling prolate trochoids. When the sampling volume moves in phase with the waves (as in most cases considered below), its trajectory through the phase space remains entirely within the regions of positive horizontal velocity anomalies (Fig. 1c). In other words, the measured velocities are consistently biased relative to the Eulerian velocities at the nominal sampling location. The magnitude of the bias, however, depends on the specifics of the sampling volume's trajectory.

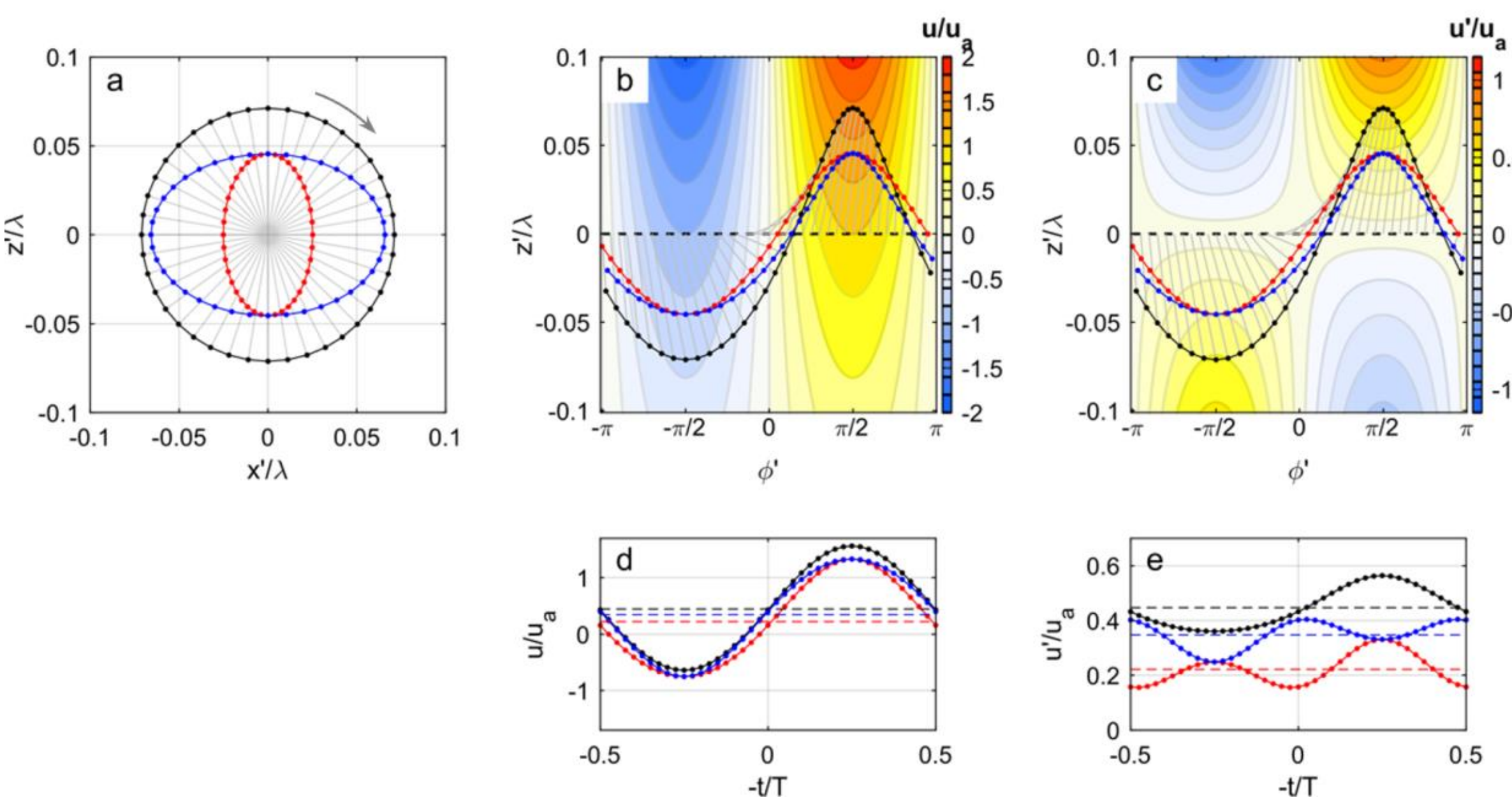

*Fig. 1 Origins of wave-induced biases. Three representative sampling volume trajectories are shown in (a) physical space, and (b, c) wave phase space. Background contour plots show (b) horizontal component of wave orbital velocities $u$ and (c) their anomalies $u' = u - u|_{z'=0}$. Time series of (d) $u$ and (e) $u'$ measured along these trajectories have different non-zero averages (dashed horizontal lines). Horizontal and vertical coordinates $(x', z')$ are relative to the nominal sampling location $(x_1, z_1)$ and normalized by wavelength $\lambda = 2\pi k^{-1}$; time is normalized by wave period $T = 2\pi\omega^{-1}$ (note that the time axis is reversed to align with the phase axis $\phi' = kx' - \omega t$). Velocities are normalized by orbital velocity amplitude $u_a = a\omega e^{kz_1}$.*

Mathematically, the origins of the wave-induced biases can be understood from equation (8) that provides a linear approximation for the local variability of wave orbital velocity fields. Applying this approximation in the vicinity of the nominal sampling location $X_1 = (x_1, z_1)$, we obtain a linearized expression for the velocity sampled along an arbitrary trajectory $X_m(t)$,

$$U_m \equiv U(X_m, t) \approx U_1 + \omega k X_1' X_m'^*, \quad (9)$$

where $U_1 = U(X_1, t)$ is the Eulerian velocity at the nominal measurement location $X_1$, and $X_m' = X_m - X_1$ are the wave-induced perturbations of the sampling location[1]. In most cases considered here $X_m'$ is partially coherent and in phase with the wave orbital motions $X_1'$, i.e. both motions include components proportional to $e^{i\phi}$. Because of that, the quadratic term $X_1' X_m'^*$ has a component proportional to $e^{i\phi} e^{-i\phi} = 1$ that leads to a non-periodic wave-induced bias. It can be readily seen that such a bias would scale with the amplitudes of both the wave and the sampling volume motions. Re-writing (9) in components,

$$U_m \approx U_1 + \omega k (x_1' + i z_1')(x_m' - i z_m') = U_1 + \omega k [(x_1' x_m' + z_1' z_m') + i(-x_1' z_m' + z_1' x_m')], \quad (10)$$

demonstrates that both the horizontal and vertical components of orbital motions contribute to the horizontal velocity bias. The vertical velocity bias vanishes when averaged over the wave period as long as the motion of the sampling volume is in phase with the wave but can be non-zero if there is a phase shift.

Next, we will explore the emergence of wave-induced biases for various platform motions.

### *3.2. Platform motion*

We consider a quasi-Lagrangian platform conducting velocity measurements remotely within a sampling volume some distance $r$ away along the vertical platform axis (Fig. 2). Our analysis is equally applicable to subsurface upward-looking (Fig. 2a) and surface downward-looking (Fig. 2b) measurements; we will demonstrate these cases in parallel, since we anticipate that the reader will be mostly interested in one or the other. For now, we assume that the platform axis remains vertical (this assumption will be relaxed in the next section). Furthermore, we are not yet concerned with the fact that actual ADCPs measure velocities using a spreading fan of several acoustic beams – these are non-trivial and will be considered separately in section 4. Instead, we assume that the vector velocity measurements can be sampled remotely and accurately. To our knowledge, such remote vector velocity sampling is currently not technologically feasible. Although convergent-beam and bistatic Doppler sonar systems do exist (e.g., Harding et al. 2021), they are not likely to be implemented on a mobile platform. Perhaps the closest approximation of a remote vector sampler is a velocimeter located some distance down the mooring line responding to the motion of the float, as discussed originally by Pollard (1973). Nonetheless, consideration of the "vector" sampling provides valuable insight into the origins of the wave-induced biases.

[1] The form of the equation (9) is not immediately intuitive. There seems to be no obvious physical reason why the sampled velocity anomaly should scale with the product of two trajectories, one of which is conjugated. This expression appears to be a fortunate consequence of the particular relationship (7) between the velocity gradients and orbital motions that holds for linear surface waves, though it is not universally applicable. Nevertheless, eq. (9) proves to be both useful and insightful.

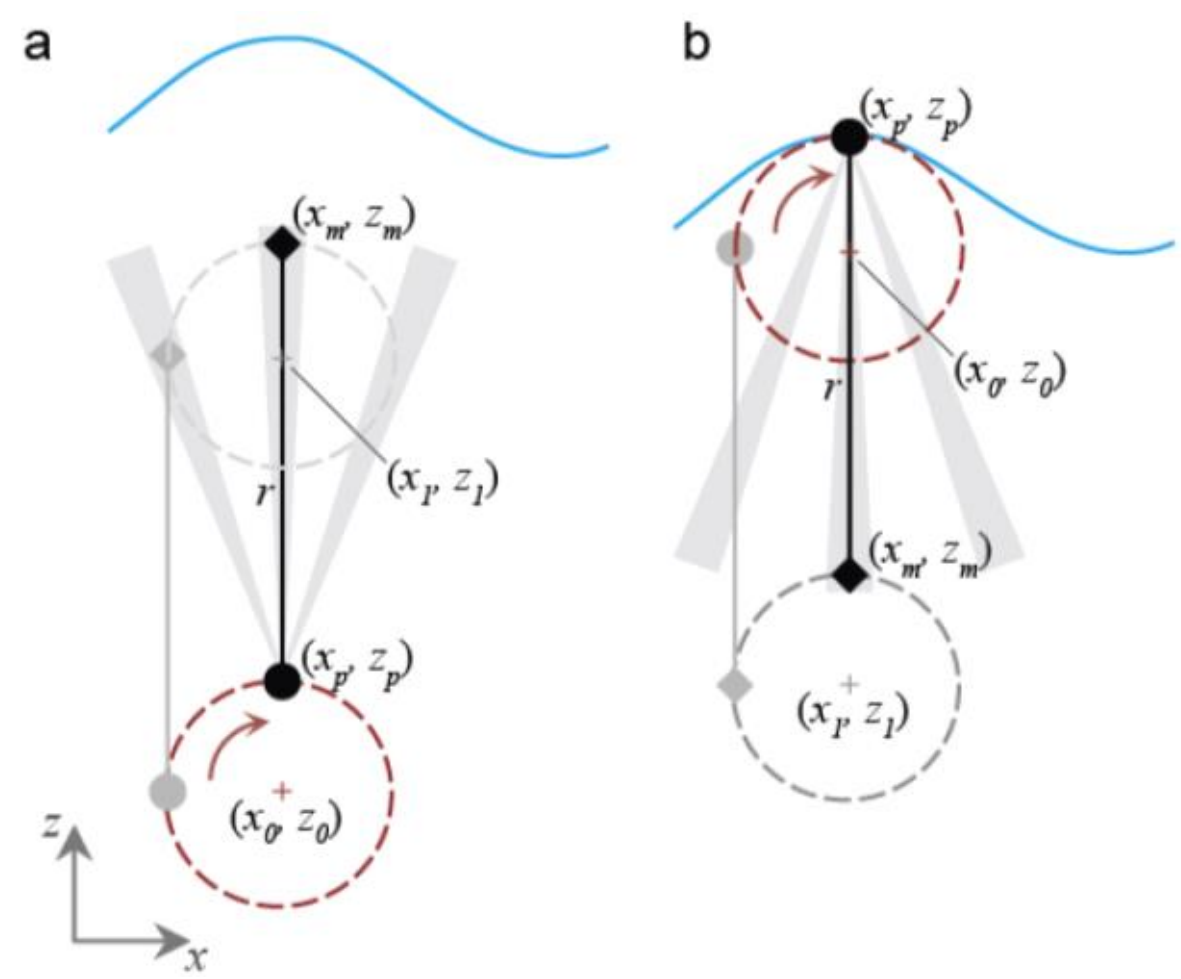

*Fig. 2 Geometry of the velocity measurements from quasi-Lagrangian platforms affected by wave orbital motions in the $x - z$ plane: a) a subsurface platform measuring velocities at a range $r$ directly above; b) a surface platform measuring velocities at a range $r$ directly below. Trajectories of the platform $(\bullet) = (x_p, z_p)$ and the sampling volume $(\blacklozenge) = (x_m, z_m)$ are shown by red and grey dashed lines. "Vector" velocity sampling (i.e., disregarding ADCP beam spread effects) is considered. The sea surface is shown in blue for reference (surface elevation is not to scale).*

At the core of the analysis is the proposition that the platform follows a fluid particle's orbital motions (3) at its nominal location $X_0$ (whether on the surface or below). The platform velocity is given by the linear expansion of the velocity field around $X_0$, eq. (8), by setting $\Xi' = X_0{}'$:

$$U_p = U_0 + \omega k X_0' X_0'^* = U_0 + a\omega k e^{i\phi_0 + kz_0} \cdot a e^{-i\phi_0 + kz_0} = U_0 + U_S, \qquad (11)$$

where

$$U_S = a^2 \omega k e^{2kz_0} \tag{12}$$

is the Stokes drift. The fluid particle trajectory that positive results in Stokes drift is illustrated by the black trace in the wave phase space diagrams (Fig. 1).

As a result of the platform motion, the sampling volume describes a similar trajectory, but around a different nominal measurement center $X_1 = \mathrm{X}_0 + ir$, where $r = z_1 - z_0$ is the nominal vertical range to the sampling volume, for an upward-looking instrument. In this section, the platform remains upright, so the sampling volume trajectory is simply[2]

$$X_m = X_1 + X_0'. \tag{13}$$

Using the linearized expression for the absolute velocity in the sampling volume (9), and noticing that in this case $X_m' \equiv X_m - X_1 = X_0'$, we obtain

$$U_m \equiv U(X_m, t) \approx U_1 + \omega k X_1' X_0'^* \,, \tag{14}$$

where $U_1 = U(X_1, t)$ is the Eulerian velocity at the nominal measurement location $X_1$. Here, it is worth noticing the similarities and differences between expression for the Stokes drift, (8), and that for the sampled velocity (14): In both cases, aperiodic contributions (i.e., biases) arise from quadratic terms of the form $X'X'^*$, except that in the latter case the product involves two different but coherent trajectories – one corresponding to the wave orbitals at the nominal measurement location (${X_1}'$) and the other to the platform motion ($X_m' = {X_0}'$).

Substituting the expression for the platform trajectory (3), we get

$$U_m = U_1 + a\omega k e^{i\phi_1 + kz_1} \, a e^{-i\phi_0 + kz_0}. \tag{15}$$

If the sampling volume is directly above or below the platform (i.e., $x_1 = x_0$, and therefore $\phi_1 = \phi_0$), this expression simplifies to

$$U_m = U_1 + a^2 \omega k \; e^{k(z_1 + z_0)} = U_1 + U_w. \tag{16}$$

Therefore, the motion of the instrument introduces an aperiodic bias

$$U_w = a^2 \omega k \; e^{k(z_1 + z_0)} \tag{17}$$

relative to the Eulerian velocity at the nominal measurement location. Vertical and horizontal components of wave-induced platform motion contribute equally to the bias in this case, as can be seen from (10).

[2] In terms of notation, the subscript "0" will always denote to the wave's orbital properties (excursions, velocities) at the platform depth, "1" will denote the same properties at the nominal sampling depth. The subscripts "*p*" and "*m*" will denote the actual motion of the platform and the "measurement" (sampling volume), respectively (see illustration in Fig. 2). Other subscripts will be defined as needed.

Expression (17) for the wave-induced bias is a little more general than that obtained by Pollard (1973) and written in a form that is better suited for interpretation of ADCP measurements from an untethered platform at arbitrary depth. Normalized profiles of wave-induced bias in absolute horizontal velocity are shown in Fig. 3–4 for a surface and subsurface instrument.

If $z_1 = z_0$, the wave-induced bias reduces to Stokes drift (12), which, again, highlights the inherent similarity of the underlying kinematics. For arbitrary vertical separation between the instrument and the sampling volume, the bias, which can be called "pseudo-Stokes drift," generally scales with $U_S$, but with a modified vertical profile (Fig. 3). This can be emphasized by re-writing (17) as

$$U_w = U_{S0} e^{k(z_1 + z_0)}, \tag{18}$$

where $U_{S0} = a^2 \omega k$ is the surface Stokes drift. Note that the wave-induced bias is purely horizontal, just as the Stokes drift.

Another potentially illuminating expression for the wave-induced bias is as the geometric mean of the Stokes drift at the nominal levels of the instrument and the measurement,

$$U_w = \left(U_S(z_0) U_S(z_1)\right)^{½}. \tag{19}$$

A practical corollary is that for measuring velocity at a given depth $z_1$, an upward-looking instrument ($z_0 < z_1$, Fig. 3a) would produce much smaller wave-induced bias than a downward-looking one ($z_0 > z_1$, Fig. 3b), all other factors being equal. This result is intuitive, as the wave-induced motions of the platform (and thus the sampling volume) are smaller in the upward looking configuration.

It is important to emphasize that expression (17) is for the *absolute* velocity, which is rarely measured. Instead, most instruments, including ADCPs, measure *relative* velocity. To obtain the relative velocity bias, the platform's own Lagrangian speed, equal to the Stokes drift, needs to be subtracted:

$$U_{wr} \equiv U_w - U_p = U_w - U_S(z_0) = a^2 \omega k e^{kz_0} (e^{kz_1} - e^{kz_0}). \tag{20}$$

As discussed in Pizzo et al. (2023) and in Dr. Pizzo's review of this article, platform motion equations should be written in terms of Lagrangian parameters. In particular, Lagrangian frequency, Doppler-shifted due to the platform's Stokes drift, should be used. In our first-order approximation, however, these distinctions do not come into play. They will, however, become important in consideration of self-propelled platforms (section 5).

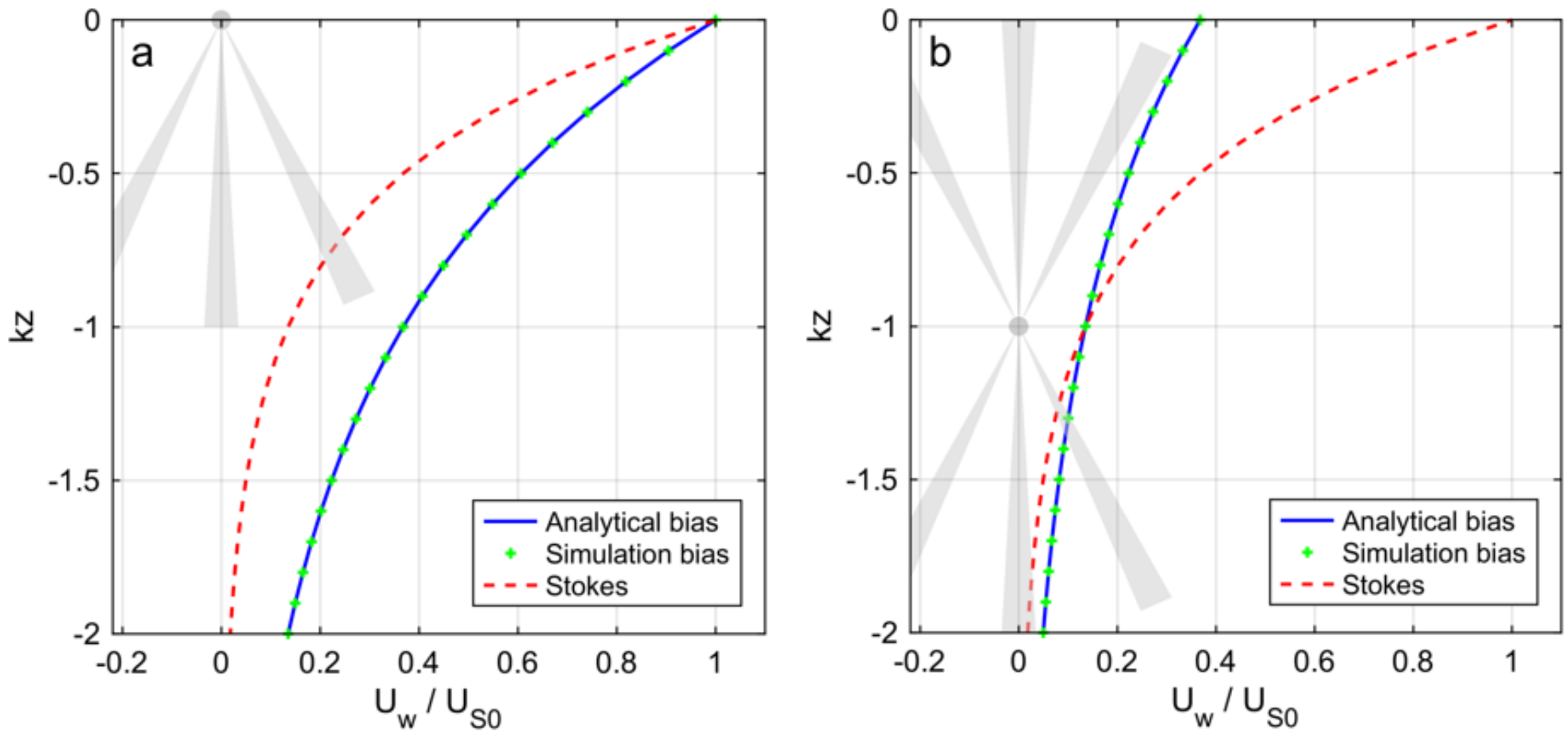

*Fig. 3 Wave-induced bias in absolute horizontal velocity measured by a) surface, downward-looking; and b) subsurface, upward- and downward-looking quasi-Lagrangian instrument. Estimates according to the analytical expression (20) are shown. Results of the semi-analytical simulation are also shown for comparison (green dots). Grey schematic shapes show the nominal depth of the instrument ($z_0 = 0$ and $z_0 = k^{-1}$, respectively). Stokes drift profiles are shown by red dashed lines, for reference. Velocity values are normalized by surface Stokes drift, $U_{S0}$; depth is normalized by the inverse wavenumber, $k^{-1}$.*

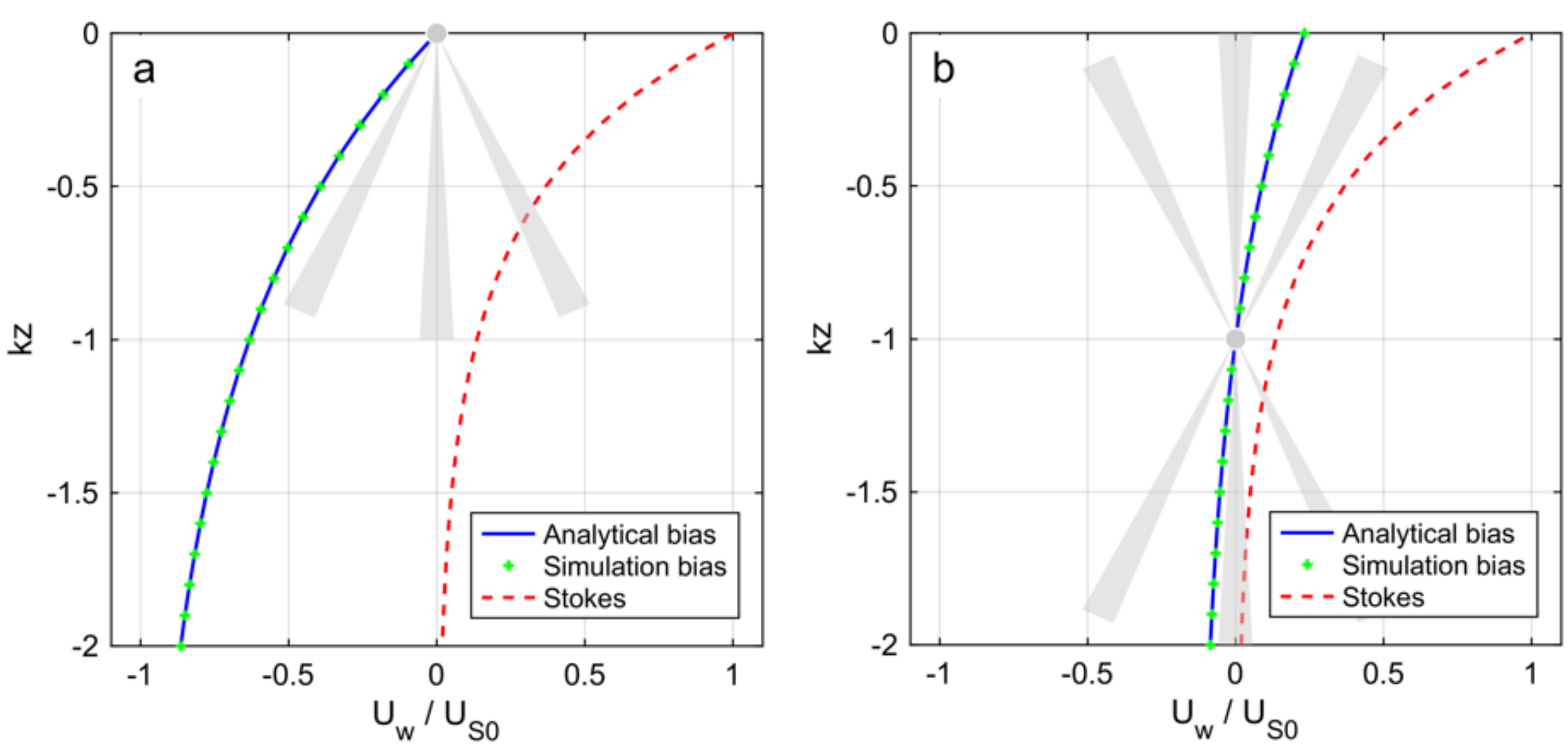

*Fig. 4 Same as Fig. 3, but for the relative velocity bias described analytically by eq. (20). Both the relative velocity and its bias vanish at the nominal instrument depth.*

### 3.3. *Platform tilting*

Wave motions are expected to affect the platform *tilt* as well, inducing additional effects on the velocity measurements. A complicating factor is that the platform tilt response to the wave forcing depends on its shape, buoyancy, and mass distribution. There are two limiting cases (Longuet-Higgins 1986): A platform with strong righting moment or a wide hull would align itself with the local effective gravity vector (Fig. 5a, the "hydrostatic" response). In contrast, a platform with a substantial vertical "keel" and a weak righting moment would be primarily tilted by the vertical shear of the horizontal orbital velocities and therefore follow the tilting of the vertical material line[3] (Fig. 5b, the "inertial" response).

The effective gravity vector is given by the vector sum of the gravitational acceleration $-ig$ and the inertial acceleration $-A$, which is the negative of the local fluid particle acceleration vector,

$$A = \frac{\partial^2 X}{\partial t^2} = -a\omega^2 e^{i\phi_0 + kz_0}. \tag{21}$$

This gives

$$G = -ig - A = -ig + a\omega^2 e^{i\phi_0 + kz_0}, \tag{22}$$

Under the small wave slope approximation, $|A| \ll g$, so the magnitude of the effective gravity vector can be estimated as

$$|G| = \sqrt{(g - a\omega^2 e^{kz_0} \sin\phi_0)^2 + (a\omega^2 e^{kz_0} \cos\phi_0)^2} \approx g - a\omega^2 e^{kz_0} \sin\phi_0. \tag{23}$$

A unit vector in the direction of effective gravity is then

$$\hat{G} \equiv \frac{G}{|G|} \approx a\omega^2 g^{-1} e^{kz_0} \cos\phi_0 - i. \tag{24}$$

(In the above derivation, the terms proportional to $\sin\phi_0$ cancel upon the expansion of the denominator.) This expression can be simplified using the deep-water dispersion relationship $\omega^2 = gk$, giving

$$\hat{G} = ake^{kz_0} \cos\phi_0 - i. \tag{25}$$

A platform with hydrostatic response would align its "mast" axis with the effective anti-gravity ("up") direction,

$$-\hat{G} = -ake^{kz_0} \cos\phi_0 + i. \tag{26}$$

Orientation of the vertical material line is given by

[3] Here, "vertical" is used to describe a material (fluid) line that is vertical *on average*. Wave-induced deformation would cause such a line to tilt periodically about the true vertical.

$$M \equiv \frac{\partial X}{\partial z_0} = \frac{\partial}{\partial z_0}\left(x_0 + iz_0 + ae^{i\phi_0 + kz_0}\right) = \; ake^{i\phi_0 + kz_0} + i, \tag{27}$$

with the corresponding unit vector

$$\widehat{M} = ake^{kz_0}\cos\phi_0 + i. \tag{28}$$

It is evident that, in case of the deep-water waves, the vectors $(-\widehat{G})$ and $\widehat{M}$ are symmetric about the vertical and 180° out of phase (this no longer holds true for finite-depth waves, see section 8).

Both modes of platform response can be generalized by expressing the platform "mast" axis orientation vector as

$$\widehat{V} = \; \gamma ake^{kz_0}\cos\phi_0 + i. \tag{29}$$

Here, the tilt response factor $\gamma = -1$ corresponds to the hydrostatic response mode ($\widehat{V} = -\widehat{G}$), $\gamma = 1$ corresponds to the inertial response ($\widehat{V} = \widehat{M}$), and $-1 < \gamma < 1$ to intermediate (attenuated) behavior of the platform. Moreover, we can later consider *complex* values of $\gamma$, possibly with $|\gamma|>1$, to represent a phase-shifted resonant response.

Tilting of the platform has two effects: a) additional distortion of the sampling volume trajectories ("sweeping"), and b) rotation of the measured velocity vectors relative to the instrument's frame of reference. The second effect may or may not be present, depending on whether the platform attitude is measured accurately and synchronously with the velocity observations, and whether this attitude is taken into account prior to velocity averaging. We therefore consider these effects separately.

*Tilt-induced motion of sampling volume ("sweeping"):* As before, consider a sampling volume at a nominal vertical distance $r$ above the platform, $X_1 = X_0 + ir$ ($r$ can be negative for a downward-looking instrument). In the presence of wave-coherent platform tilt, trajectory of this sampling volume $X_m$, is comprised of translational motion following the wave orbitals ($X'_p = X'_0$, as in the no-tilt case) and rotational ("sweeping") motion caused by the tilting of the "mast" vector, $r\widehat{V}$ (Fig. 5):

$$X_m = X_0 + X'_0 + r\widehat{V}. \tag{30}$$

Perturbations relative to the nominal sampling volume location $X_1$ is therefore

$$\begin{aligned} X'_m \equiv X_m - X_1 &= a\,e^{i\phi_0 + kz_0} + \gamma akr\,e^{kz_0}\cos\phi_0 \\ &= a\,e^{kz_0}[(1+\gamma kr)\cos\phi_0 + i\sin\phi_0]. \end{aligned} \tag{31}$$

As illustrated in Fig. 5a, the hydrostatic platform response ($\gamma = -1$) causes retrograde[4] tilt of the instrument axis and reduces the horizontal excursion of the sampling volume at shorter ranges above the instrument ($0 < kr < 2$). Conversely, the horizontal excursions are amplified at longer ranges ($kr > 1$) and below the instrument ($kr < 0$). For the inertial platform response ($\gamma = 1$), the platform axis tilt is prograde, and the sampling volume trajectory is distorted in the opposite sense (Fig. 5b). The semi-analytical model shows similar distortion of the sampling volume trajectories (Fig. 6, central beam, disregard the slanted beams for now), but the shapes deviate slightly from true ellipses due to higher-order terms in the platform axis vector $\hat{V}$. It can be anticipated that enhanced amplitude of the sampling volume orbiting would lead to an increase in the wave-induced bias and vice versa.

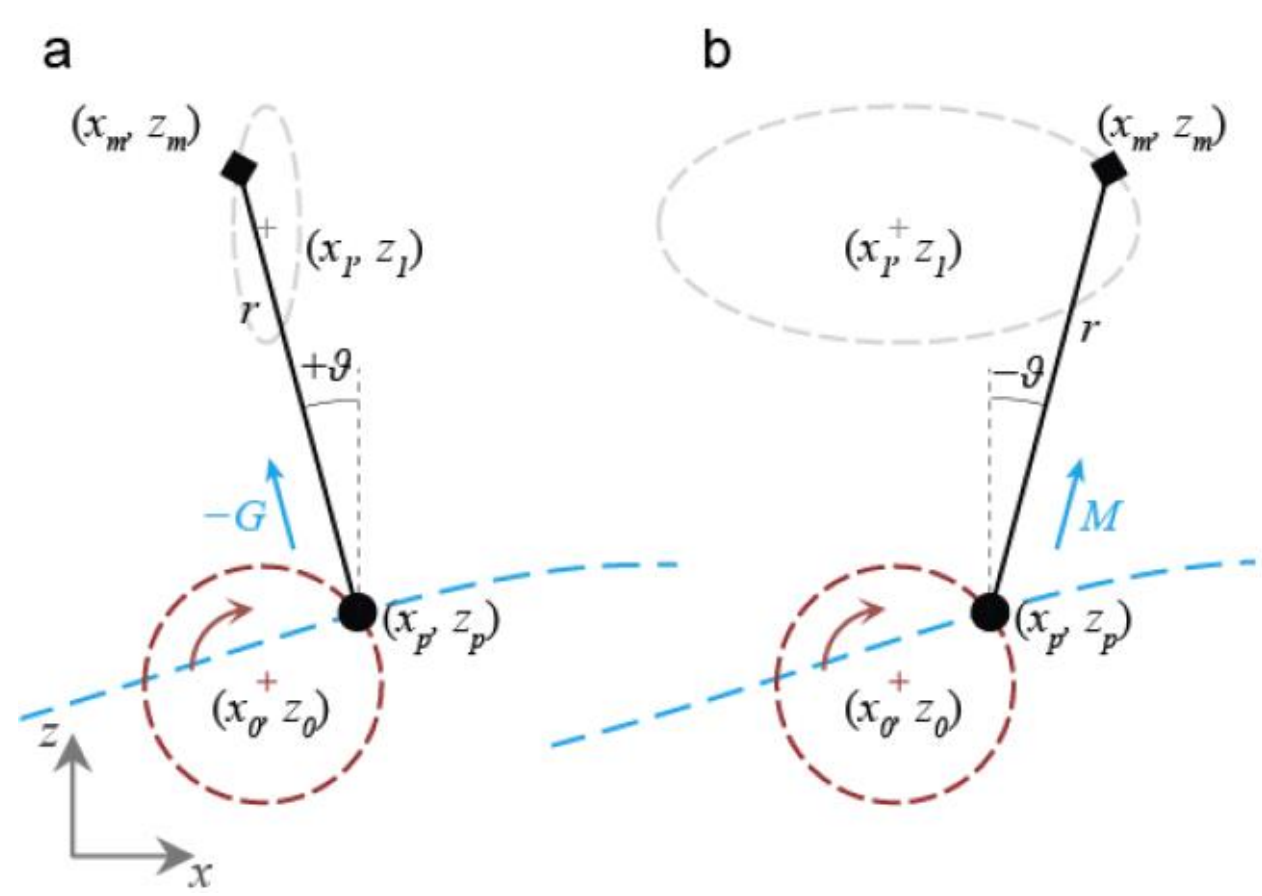

*Fig. 5 Geometry of the velocity measurements from a platform tilted by the waves. Two limiting cases of platform response are shown: a) "hydrostatic" response (platform aligns with the local anti-gravity direction, $-G$); and b) "inertial" response (platform aligns with the vertical material line vector $M$). Trajectories of the platform (●) = $(x_p, z_p)$ and the sampling volume (◆) = $(x_m, z_m)$ are shown by red and grey dashed lines. Isobaric (isopotential) surfaces are shown by blue dashed lines. Note the elliptical distortion of the sampling volume $(x_m, z_m)$ trajectories.*

[4] Here, 'prograde' / 'retrograde' refers to the vertical axis tilting in the same/opposite sense relative to the horizontal displacement.

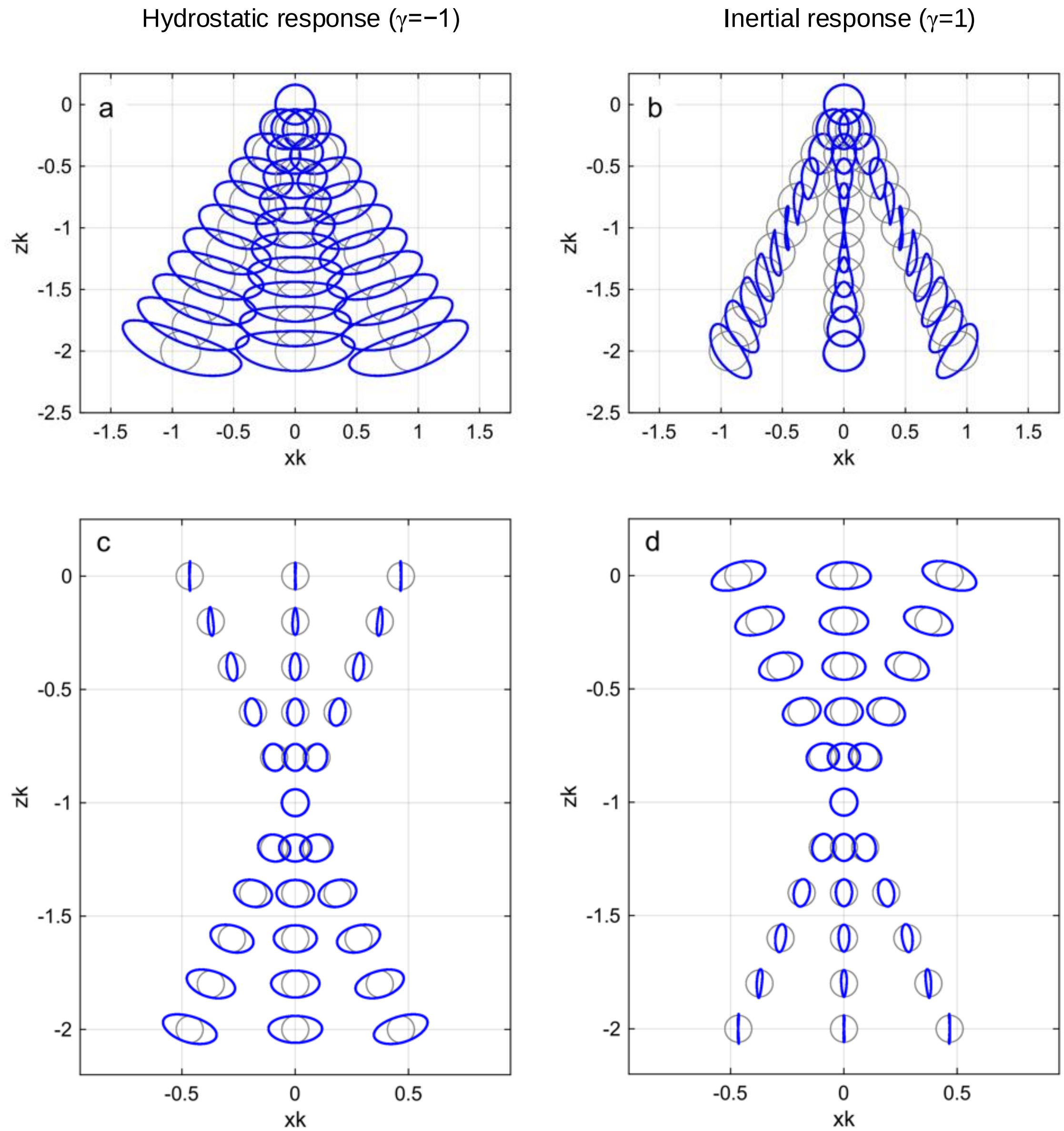

*Fig. 6 Wave-induced sampling volume trajectories (blue) estimated using a semi-analytical model. Examples of a surface (a, b) and a sub-surface (c, d) platform are shown for hydrostatic ($\gamma = -1$, left) and inertial ($\gamma = 1$, right) platform tilt response models. Trajectories in absence of the platform tilt are shown in grey, for reference. The central "beam" corresponds to a "vector" sampler; the two slanted beams represent a pair of ADCP beams at 25° from the vertical. Axes are normalized by inverse wavenumber $k^{-1}$.*

Similarly to (14), absolute Eulerian velocity at the measurement location is given by

$$\begin{aligned} U_m &= U_1 + \omega k X_1{}' X_m'^{*} = \\ &= U_1 + a\omega k e^{i\phi_1 + kz_1} a\, e^{kz_0}\left(e^{-i\phi_0} + \gamma kr\, e^{i\phi_0}\cos\phi_0\right) = \\ &= U_1 + a^2\omega k e^{k(z_1+z_0)}\left(e^{i(\phi_1-\phi_0)} + \gamma kr\, e^{i\phi_0}\cos\phi_0\right) = \\ &= U_1 + U_w + U_t. \end{aligned} \tag{32}$$

In addition to the aperiodic wave-induced pseudo-Stokes bias $U_w$, we now have a "sweeping" term,

$$\begin{aligned} U_t{}' &= \gamma a^2 \omega k^2 r\, e^{k(z_1+z_0)}\, e^{i\phi_0}\cos\phi_0 = \\ &= \gamma a^2\omega k^2 r\, e^{k(z_1+z_0)}\,[\cos^2\phi_0 + i\sin\phi_0\cos\phi_0]. \end{aligned} \tag{33}$$

Although $U_t{}'$ is periodic, it is not harmonic, and therefore contributes to an additional phase-averaged bias

$$\overline{U_t} \equiv \langle U_t\rangle = \tfrac{1}{2}\gamma a^2\omega k^2 r\, e^{k(z_1+z_0)} = \tfrac{1}{2}\gamma kr U_w. \tag{34}$$

Here, angle brackets represent averaging over the wave phase, and an overbar emphasizes the fact that the bias is a phase-averaged quantity. Note that in order for the linear approximation (32) to remain valid, we need to limit our consideration to $|r| \lesssim k^{-1}$, since otherwise the excursions of the sampling volume $X_m'$ may become excessive even for small platform tilts. Behavior of the sweeping bias at longer ranges may be best studied using a semi-analytical model. Examples of the sweeping bias $\overline{U_t}$ are shown in Fig. 7 and Fig. 8 and discussed below.

*Tilt-induced reference frame rotation:* If the instrument is not "aware" of its wave-coherent tilt, the actual velocities being measured (relative to the platform's motion, $U_{1r} = U_1 - U_p$) will be recorded in the instrument reference frame as

$$U_{1i} = \left(U_1 - U_p\right)(i\hat{V}^{-1}) = -ia\omega e^{i\phi_0}(e^{kz_1} - e^{kz_0})(1 + i\gamma ak e^{kz_0}\cos\phi_0), \tag{35}$$

where the multiplier $i\hat{V}^{-1} = i\hat{V}^{*} = 1 + i\gamma ak e^{kz_0}\cos\phi_0$ represents rotation from Earth to instrument frame. After the phase averaging in the instrument frame, a "frame tilt" bias emerges:

$$U_f \equiv \langle U_{1i}\rangle = \tfrac{1}{2}\gamma a^2\omega k e^{kz_0}(e^{kz_1} - e^{kz_0}). \tag{36}$$

Interestingly, comparing this with (20), we find that

$$U_f = \tfrac{1}{2}\gamma U_{wr}, \tag{37}$$

i.e., reference frame tilting results in an additional bias that is half of the wave-induced relative pseudo-Stokes bias $U_{wr}$, but can have either sign depending on the response mode.

Profiles of both components of tilt-induced bias for surface and subsurface instruments are shown in Fig. 7 and Fig. 8. As anticipated by the sampling volume trajectories (Fig. 6), and as evident from the above analytical expressions, the hydrostatic response ($\gamma = -1$) leads to the tilt biases $U_f$ and $U_t$ partially compensating the motion bias $U_{wr}$. The net bias is therefore generally reduced in these cases (Fig. 7a and Fig. 8a). Correspondingly, inertial platform response ($\gamma = 1$) leads to an overall increase in the bias amplitude (Fig. 7b and Fig. 8b). A resonant platform response (not shown), characterized by a complex $\gamma$ representing a phase shift of the response, is the only case among those considered here that can produce a wave-induced bias in the vertical velocity. Note also that, unlike for the other biases considered here, the sweeping bias depends on $kr = k(z_1 - z_0)$ in addition to $k(z_1 - z_0)$, and therefore the profiles shown in Fig. 7 and Fig. 8 are not universal, even when scaled as shown. Analytical expressions developed here should be used to assess the bias profiles in each particular case.

## 4. ADCP beam geometry and velocity response function

ADCP measures the radial velocities along several beams (typically $4 - 5$). From these, profiles of all velocity components can be derived (e.g., Gordon, 1996). An important assumption underlying ADCP velocity calculations is that the velocity field does not vary horizontally at scales comparable to the beam separation. This assumption, however, can be easily violated by the surface waves' orbital velocities. As will be shown next, the wave phase difference across the ADCP beams does not introduce an additional bias per se, but it may substantially modify the motion and tilt biases.

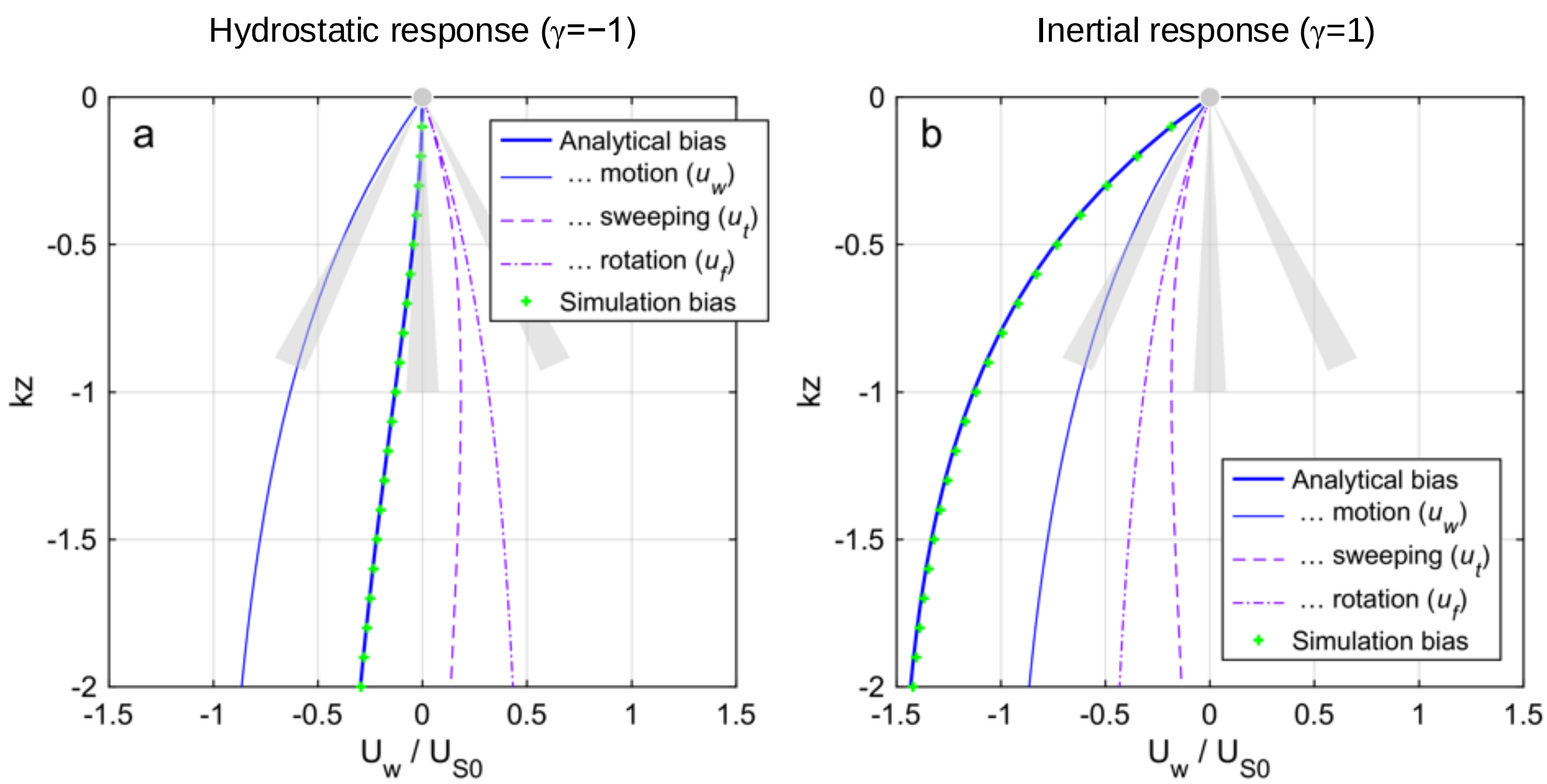

*Fig. 7 Wave-induced tilt and motion biases in relative horizontal velocity measured by a surface downward-looking Lagrangian instrument with a) hydrostatic and b) inertial response. Estimates of motion bias ($u_w$, eq. (20)), sweeping tilt bias ($u_t$, eq. (34)), and frame rotation bias ($u_f$, eq. (36)), and total bias are shown. Results of the semi-analytical simulation are shown for comparison (green dots).*

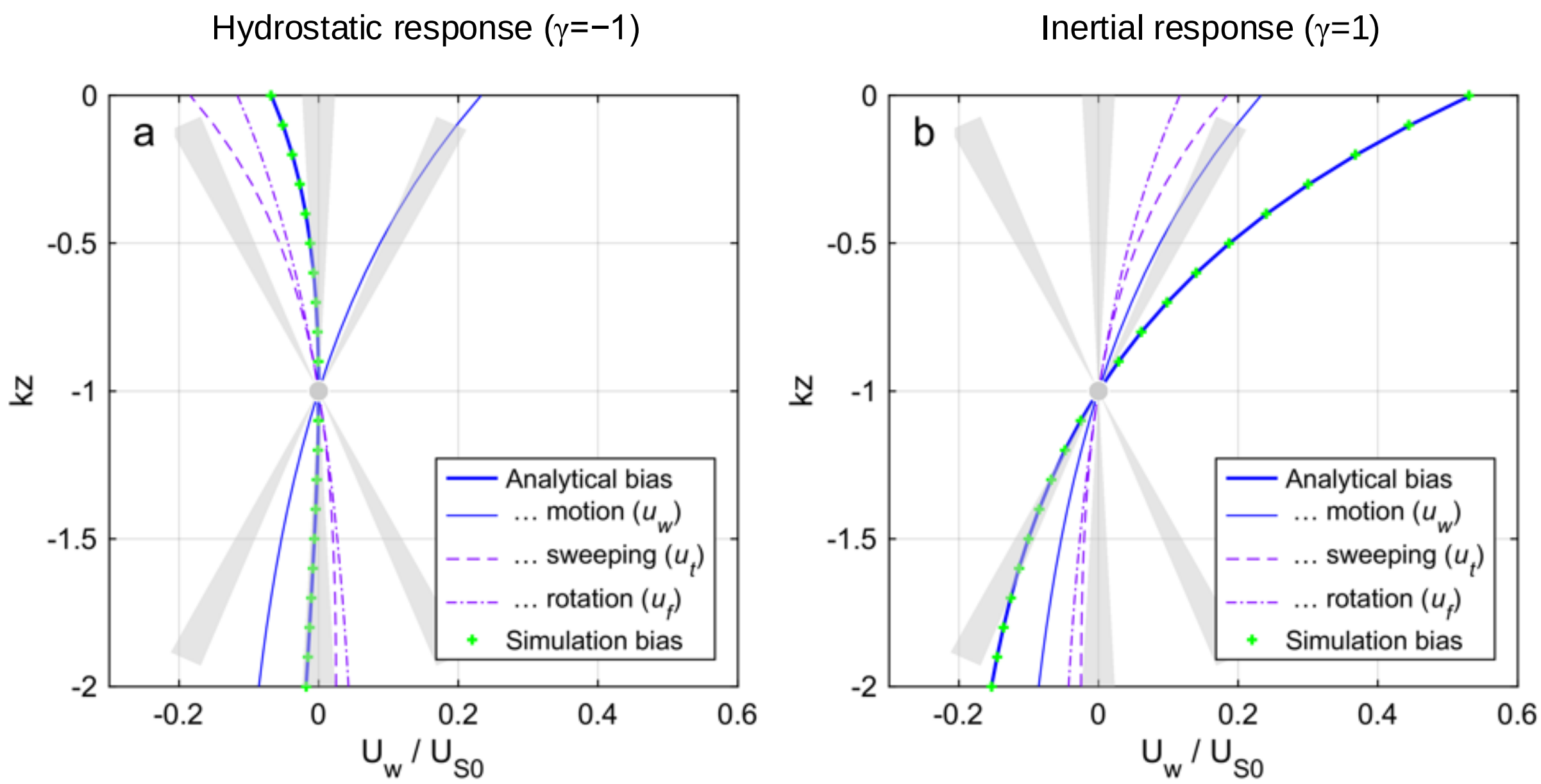

*Fig. 8 Same as Fig. 7, but for a subsurface Lagrangian instrument. Notice a factor of 3 difference in horizontal axis scale between Figs. 7 and 8.*

Generally, the effect of the transformation from the ADCP beam velocities to horizontal and vertical velocity estimates can be represented by the multiplicative response functions $R_u$ and $R_w$,

$$\begin{aligned} u_{ADCP} &= R_u u_1, \\ w_{ADCP} &= R_w w_1, \end{aligned} \tag{38}$$

where $(u_1, w_1)$ is the true velocity at the nominal measurement location. Gargett et al. (2008, 2009) provide expressions for these response functions for a general case of a "turbulent" velocity field with partially-coherent components. The plane wave considered here is a particular case of such a field, so the Gargett et al. formulas are fully applicable (with a suitable choice of phase and amplitude parameters). Nonetheless, we provide a simplified derivation of these response functions below for completeness.

Consider a pair of symmetrically slanted upwards-looking ADCP beams in the vertical $x - z$ plane aligned with the wave propagation direction. This arrangement would correspond to a pair of opposing beams on a typical 4-beam ADCP with a Janus beam pattern. The ADCP records the along-beam velocities,

$$B^{\pm} = \pm u \sin\beta + w \cos\beta, \tag{39}$$

where $\beta$ is the ADCP beam angle (typically, 20−30° depending on the model), and the "+" sign corresponds to the beam pointing towards $+x$.

In complex notation,

$$B^{\pm} = \Im[U(\cos\beta + i\sin(\pm\beta))] = \Im\left[Ue^{\pm i\beta}\right], \tag{40}$$

where $\Im(\cdot)$ is the imaginary part operator. This expression can be understood easily by noting that the $e^{\pm i\beta}$ factor rotates the vectors in such a way that the corresponding beam aligns with the vertical (imaginary) axis.

From these, the horizontal and vertical components of velocity can be derived as

$$\begin{aligned} u_{ADCP} &= \frac{B^+ - B^-}{2\sin\beta}, \\ w_{ADCP} &= \frac{B^+ + B^-}{2\cos\beta}. \end{aligned} \tag{41}$$

At the nominal measurement level, $z_1 = z_0 + r$, where $r$ is the nominal vertical range, sampling volumes of the two beams have nominal horizontal coordinates

$$x_1^{\pm} = x_0 \pm r\tan\beta. \tag{42}$$

Therefore, the wave phases $\phi_1^{\pm}$ observed by the two beams differ from $\phi_0$ and from each other:

$$\phi_1^{\pm} \equiv kx_1^{\pm} - \omega t = \phi_0 \pm kr\tan\beta. \tag{43}$$

Let's disregard the wave-induced motion of the instrument for a moment (e.g., consider a case of a bottom-mounted ADCP as in Gargett et al. (2008, 2009)). The measured beam velocities are

$$\overline{U_m^{\pm}} = a^2\omega k e^{k(z_1+z_0)} \left\langle e^{i\left(\phi_1^{\pm} - \phi_0\right)} \right\rangle = a^2\omega k e^{k(z_1+z_0)}\, e^{\pm ikr\tan\beta} \tag{44}$$

The horizontal velocity inferred from these measurements is

$$\begin{aligned} u_{ADCP} &= \frac{B^+ - B^-}{2\sin\beta} = \frac{-a\omega e^{kz_1}}{2\sin\beta}\left[\cos(\phi_1^+ + \beta) - \cos(\phi_1^- - \beta)\right] = \\ &= \frac{a\omega e^{kz_1}}{\sin\beta}\sin\left(\frac{\phi_1^+ + \phi_1^-}{2}\right)\sin\left(\beta + \frac{\phi_1^+}{2}\right) = \\ &= \frac{a\omega e^{kz_1}}{\sin\beta}\sin(\phi_0)\sin(\beta + kr\tan\beta) = u_1\frac{\sin(\beta + kr\tan\beta)}{\sin\beta}, \end{aligned} \tag{45}$$

where $u_1 = a\omega e^{kz_1}\sin(\phi_0)$ is the wave orbital velocity at the nominal measurement location.

Therefore, the inferred velocity differs from the actual wave orbital velocity by a response function factor

$$R_u \equiv \frac{u_{ADCP}}{u_1} = \frac{\sin(\beta + kr\tan\beta)}{\sin\beta}. \quad (46)$$

This expression is equivalent to the square root of the variance response function presented by Gargett et al., (2009, eq. 13) with the assumption of a 1:1 aspect ratio and a 90° phase difference between the $u$- and $w$-components of the velocity field, as required for the plane wave considered here. Similarly,

$$R_w \equiv \frac{w_{ADCP}}{w} = \frac{\cos(\beta + kr\tan\beta)}{\cos\beta}. \quad (47)$$

It is important to note that even though expressions for the response function (46) and (47) were derived for an upward-looking ADCP ($r > 0$), they remain valid for a downward-looking instrument ($r < 0$). This can be demonstrated by repeating the calculations after making the appropriate changes to the expressions (40) and (42).

The response functions (46) and (47) have several peculiarities. We illustrate their behavior by showing their dependence on normalized range $kr$ for beam angles 20−30° in Fig. 9.

One of the key results is that the ADCP response is inherently *asymmetric*, i.e., the response functions for upward- and downward-looking instruments are different at the same range. This asymmetry is somewhat unexpected for such a symmetric measurement system as an ADCP. The origins of this asymmetry can be traced in the illustration shown in Fig. 10. The horizontal separation between the sampling volumes of the two beams causes a change of wave phase and an additional rotation of the wave orbital velocity vector relative to that at the nominal measurement location (0, 0). For an upward-looking instrument, this rotation is opposite to that of the beam vectors (with respect to the vertical). As a result, the beam velocities are amplified relative to the no-rotation case. For a downward-looking instrument, the wave velocity vector rotates in the same sense as the beam vectors, so the apparent beam velocities are reduced. This asymmetric relationship between upward- and downward-looking ADCP measurements holds regardless of the direction of the wave propagation, as can be readily demonstrated by choosing a coordinate system aligned with the wave propagation direction. The actual cause of the asymmetry is the inherent structure of the spatial distribution of orbital velocity phase, which, in turn, is set by the downward direction of the gravitational restoring force. As a result, upward-looking ADCPs are generally at a disadvantage, as their beam configuration makes them more prone to contamination by wave orbital motions (other factors being equal).

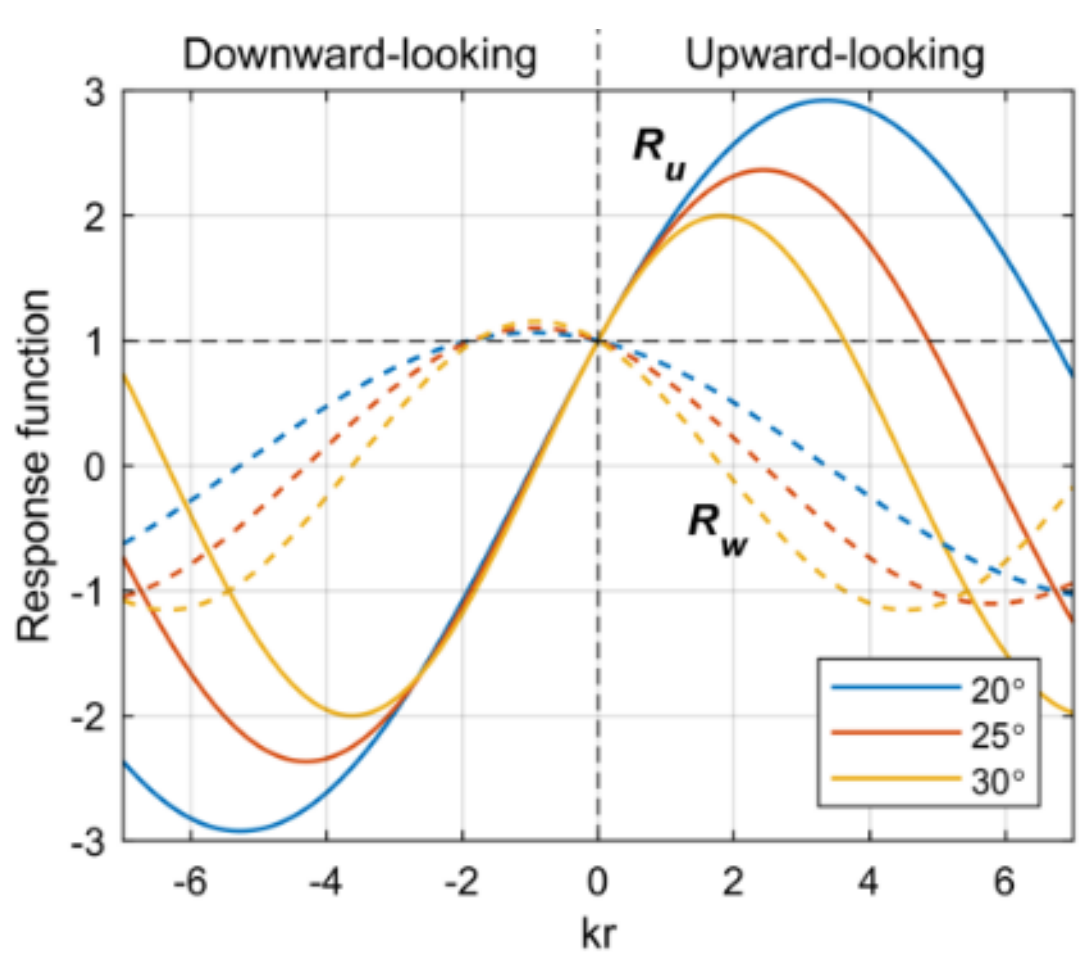

*Fig. 9 ADCP response functions for horizontal ($R_u$, solid lines) and vertical ($R_w$, dashed lines) velocities as a function of normalized range (or wavenumber), $kr$. Response functions are shown for several typical ADCP beam angles. Positive values of $kr$ correspond to upward-looking ADCPs.*

A few other peculiarities of the ADCP response functions are:

- They do not depend on the depth, only the horizontal beam separation normalized by the wavenumber.

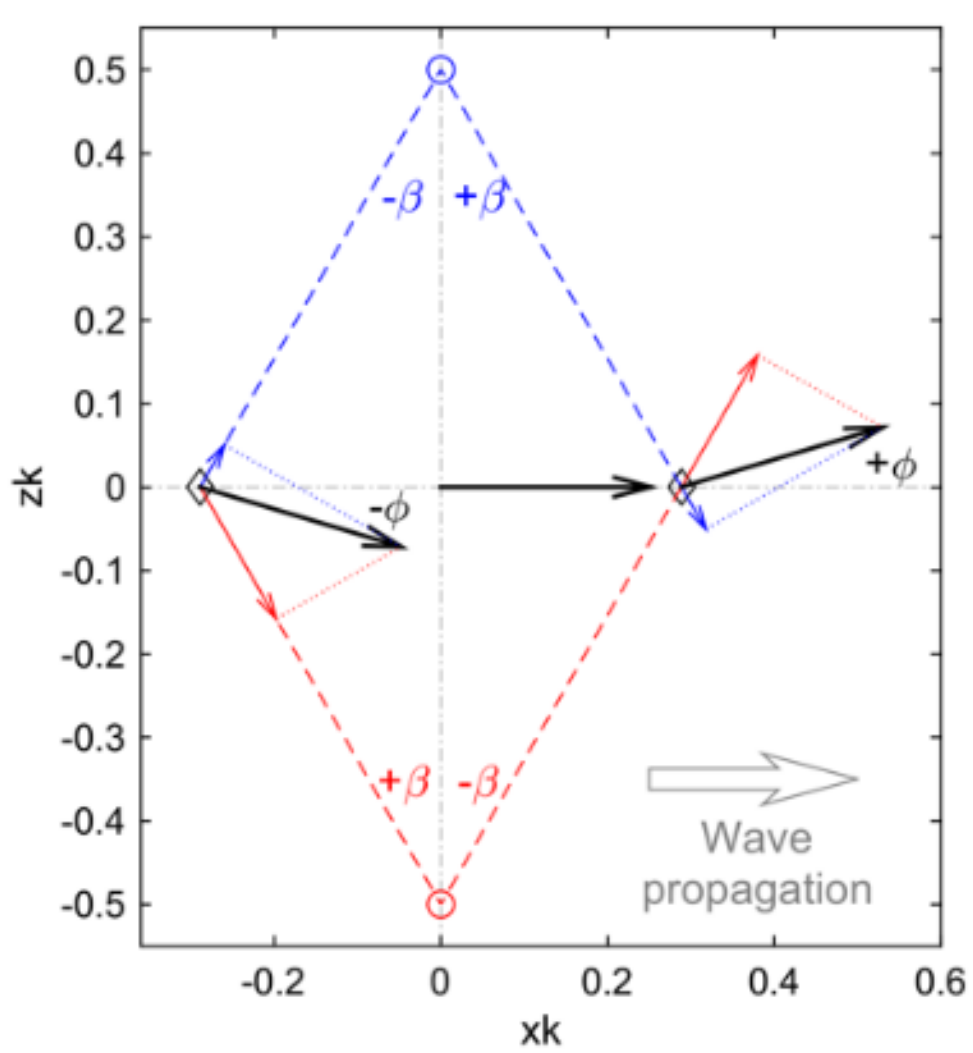

*Fig. 10 Illustration of the differences between the upward- and downward-looking ADCP response functions when sampling the same wave field. Wave field velocity vectors (black arrows) rotate clockwise with time and counterclockwise with increasing x. ADCP beams (dashed) and measured beam velocities (vectors) are shown in red for an upward-looking instrument and in blue for a downward-looking one. Both instruments have a nominal sampling volume at (0,0). The phase of the maximum horizontal velocity is shown. It is evident that the upward-looking instrument measures larger beam velocities, and therefore overestimates the horizontal orbital wave velocity.*

- They change signs and have an infinite number of zeros. The first zero of $R_u$ is achieved for $r \approx -0.9k^{-1}$ (downward-looking ADCP).
- They do not have asymptotes for long ranges (or short waves), $|kr| \to \infty$.

(Some of these peculiarities have been previously discussed by Theriault (1986), who considered ADCP response in a similar manner, but disregarded vertical velocities.)

The ADCP response function for the horizontal component of orbital velocities can reach absolute values of 2 – 3, especially for narrower beam angles (Fig. 9). In contrast, the vertical component of orbital velocity is almost always underestimated ($|R_w| \lesssim 1$). Response functions for horizontal and vertical components behave differently with increasing $kr$, so the orbital trajectories are always distorted to some extent.

Since the response functions are multiplicative, beam geometry in itself does not introduce additional biases. However, these response functions modify the wave-induced biases discussed in sections 3.1 – 3.3.

To illustrate this effect, consider how the ADCP response function affects the platform motion bias. Trajectories of the sampling volumes for the two beams are

$$X_m^{\pm} = X_0 + X_0' + ir \pm r \tan\beta, \tag{48}$$

i.e. small orbital motions $X_m^{\pm}{}' = X_0{}'$ with respect to the nominal sampling locations of the two beams, $X_1^{\pm} = X_0 + ir \pm r \tan\beta$, see Fig. 6 for an illustration.

Similarly to (15), the Eulerian orbital velocities at these locations can be expressed as

$$U_m^{\pm} = U_1^{\pm} + a^2 \omega k \, e^{i(\phi_1^{\pm} - \phi_0) + k(z_1 + z_0)}. \tag{49}$$

Since we are interested in the mean wave-induced velocity bias, we can take the phase average of $U_m^{\pm}$ now, before the beam velocity transformations (since the latter are linear). This allows us to drop the periodic first term and obtain

$$\overline{U_m^{\pm}} = a^2 \omega k e^{k(z_1 + z_0)} \left\langle e^{i(\phi_1^{\pm} - \phi_0)} \right\rangle = a^2 \omega k e^{k(z_1 + z_0)} \, e^{\pm ikr \tan\beta}. \tag{50}$$

The corresponding biases in the along-beam velocities are

$$\begin{aligned} \overline{B^{\pm}} = \Im\left[\overline{U_m^{\pm}} e^{\pm i\beta}\right] = a^2 \omega k e^{k(z_1 + z_0)} \, \Im\left[e^{\pm i(\beta + kr \tan\beta)}\right] = \\ = \pm a^2 \omega k e^{k(z_1 + z_0)} \sin(\beta + kr \tan\beta) \, . \end{aligned} \tag{51}$$

The inferred horizontal velocity bias is then

$$U_{w,ADCP} = \frac{\overline{B^+} - \overline{B^-}}{2 \sin\beta} = a^2 \omega k e^{k(z_1 + z_0)} \frac{\sin(\beta + kr \tan\beta)}{\sin\beta} = U_w R_u. \tag{52}$$

Therefore, the same response function $R_u$ modifies the wave-induced bias $U_w$.

It is important to remember that the platform's own velocity and its Stokes drift are not affected by the ADCP beam geometry. Therefore, the *relative* wave-induced bias, corrected for the ADCP response function, becomes

$$U_{wr,ADCP} = R_u U_w - U_S(z_0) = a^2 \omega k e^{kz_0}(R_u e^{kz_1} - e^{kz_0}). \quad (53)$$

Similarly, reference frame rotation bias becomes

$$U_{f,ADCP} = \frac{1}{2}\gamma a^2 \omega k e^{kz_0}(R_u e^{kz_1} - e^{kz_0}). \quad (54)$$

The effect of the beam geometry on the "sweeping" bias (section 3.3) is less straightforward. Similarly to (30), trajectories of each beam's sampling volumes due to the "sweeping" motions are

$$X_m^{\pm} = X_0 + r\hat{V} \pm r\tan\beta\left(-i\hat{V}\right) = X_0 + r(\cos\beta)^{-1} e^{\mp i\beta}\hat{V}. \quad (55)$$

Recall that the nominal sampling locations of the two beams $X_1^{\pm} = X_0 + ir \pm r\tan\beta = X_0 + ir(\cos\beta)^{-1}e^{\mp\beta}$, and therefore

$$X_m^{\pm\prime} \equiv X_m^{\pm} - X_1^{\pm} = r(\cos\beta)^{-1}e^{\mp i\beta}\big(\hat{V} - i\big) = r(\cos\beta)^{-1}e^{\mp i\beta}\,\gamma a k e^{kz_0}\cos\phi_0 \quad (56)$$

In contrast to (31), trajectories now include non-negligible vertical excursions arising from the relatively large beam angle $\beta$. Similarly to (15), the Eulerian orbital velocities at these locations can be expressed as

$$\begin{aligned} U_m^{\pm} &= U_1^{\pm} + a\omega k e^{i\phi_1^{\pm} + kz_1}\big(r(\cos\beta)^{-1}e^{\pm i\beta}\,\gamma a k e^{kz_0}\cos\phi_0\big) \\ &= U_1^{\pm} + \gamma a^2 \omega k^2 r(\cos\beta)^{-1} e^{k(z_0+z_1)}\cos\phi_0\, e^{\pm i\beta + i\phi_1^{\pm}}. \end{aligned} \quad (57)$$

Again, applying phase averaging that eliminates the periodic first term ($U_1^{\pm}$), we obtain

$$\begin{aligned} \overline{U_m^{\pm}} &= \gamma a^2 \omega k^2 r(\cos\beta)^{-1} e^{k(z_0+z_1)} e^{\pm i\beta}\left\langle \cos\phi_0\, e^{i\phi_1^{\pm}}\right\rangle = \\ &= \frac{1}{2}\gamma a^2 \omega k^2 r(\cos\beta)^{-1} e^{k(z_0+z_1)} e^{\pm i(\beta + kr\tan\beta)} = \\ &= \overline{U_t}(\cos\beta)^{-1} e^{\pm i(\beta + kr\tan\beta)}. \end{aligned} \quad (58)$$

To first order, we can disregard the change in the beam angles due to the instrument tilt, so the phased-averaged beam velocities are

$$\begin{aligned} \overline{B^{\pm}} &= \Im\left[\overline{U_m^{\pm}}e^{\pm i\beta}\right] = \overline{U_t}(\cos\beta)^{-1}\Im\left[e^{\pm i(2\beta + kr\tan\beta)}\right] = \\ &= \pm\overline{U_t}\sin(2\beta + kr\tan\beta)\,(\cos\beta)^{-1}. \end{aligned} \quad (59)$$

The inferred horizontal velocity bias due to the "sweeping" motion is

$$U_{t,ADCP} = \frac{\overline{B^+} - \overline{B^-}}{2\sin\beta} = \frac{2\sin(2\beta + kr\tan\beta)}{\sin 2\beta}\,\overline{U_t}. \quad (60)$$

The effective response function to the “sweeping” motion is therefore

$$R_t = \frac{2\sin(2\beta + kr\tan\beta)}{\sin 2\beta}. \tag{61}$$

The dependence of this response function on normalized range is shown in Fig. 11.

Interestingly, it can be seen that $R_t = R_u + R_w$, although it is not clear why it should be the case. At short ranges ($kr \to 0$), $R_t \approx 2$, which means that the ADCP processing doubles the horizontal velocity bias arising from the “sweeping” motion (although the bias vanishes at $kr = 0$).

Figures 12 and 13 show examples of wave-induced tilt and motion biases exacerbated by the ADCP response functions. The displayed profiles do not fully represent the range of situations that may arise. As shown earlier, the biases are anticipated to have a wide range of shapes depending on wave field properties, platform depth, and the ADCP beam configuration. Nevertheless, these examples clearly demonstrate that the ADCP beam geometry and the resultant velocity response functions may lead to complex and significant (a factor of ±2−3) modification of both the resolved orbital wave motions and the wave-induced biases. Therefore, consideration of the beam geometry effects is essential for accurate characterization of uncertainty in any ADCP measurements in the presence of waves.

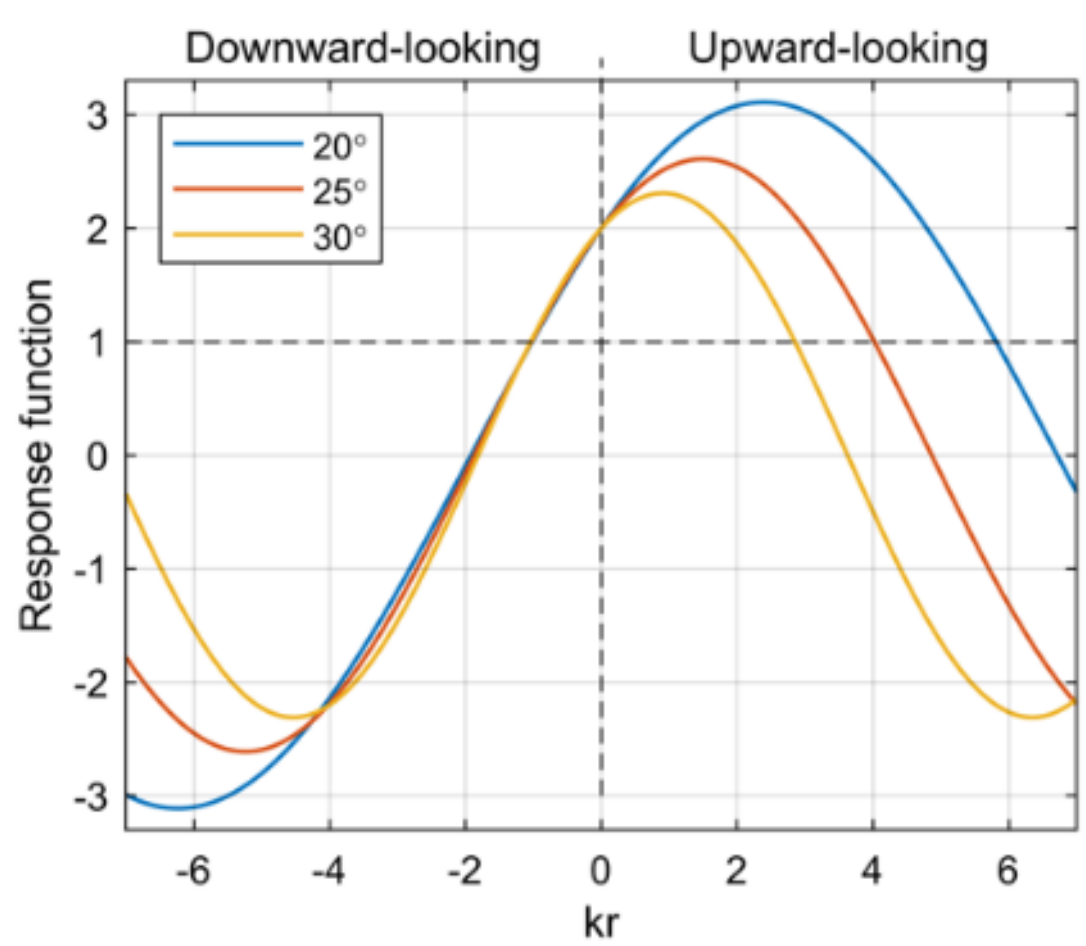

*Fig. 11 ADCP response functions for “sweeping” horizontal velocity bias $R_t$ as a function of normalized range (or wavenumber), $kr$. Response functions are shown for several typical ADCP beam angles. Positive values of $kr$ correspond to upward-looking ADCPs.*

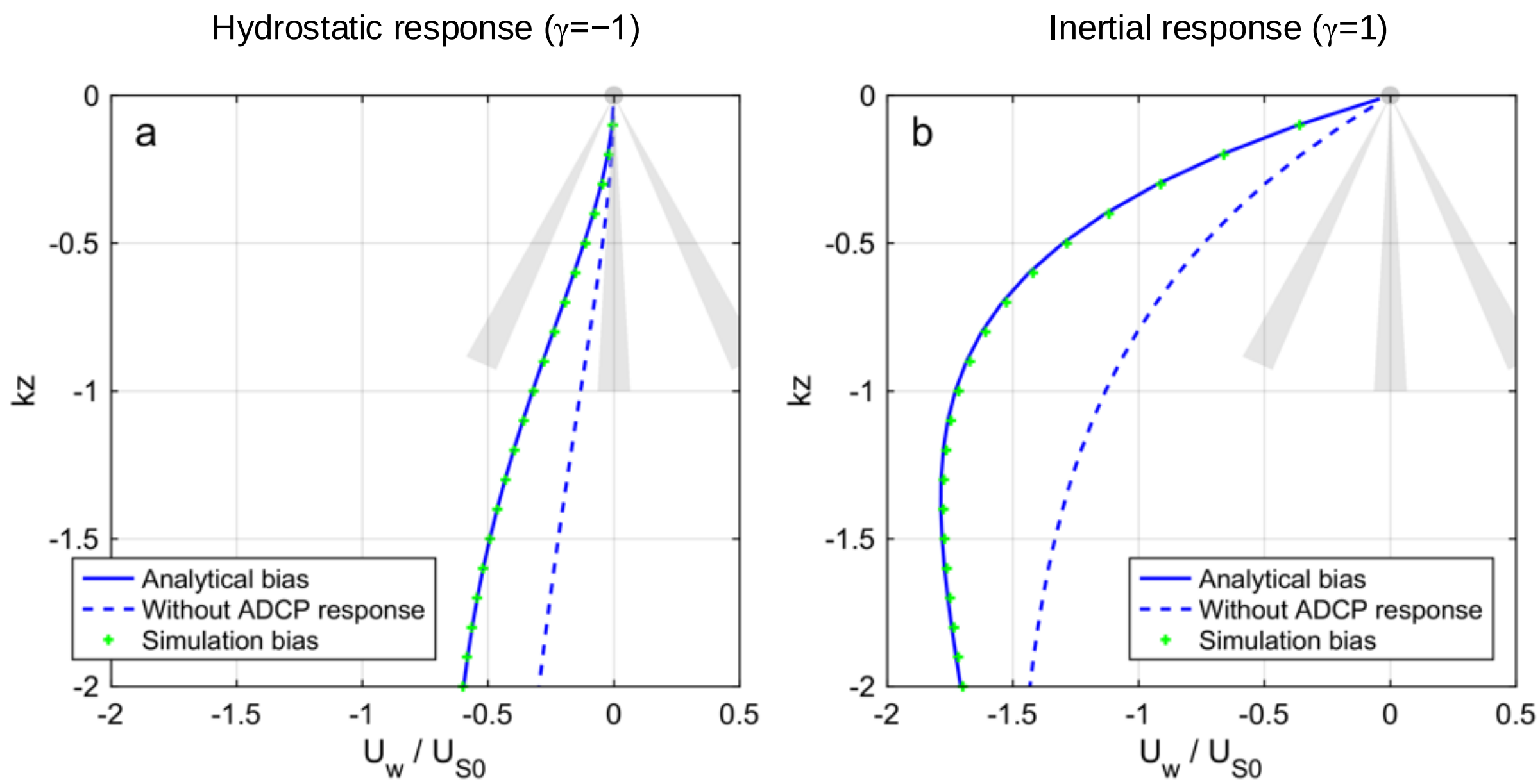

*Fig. 12 Wave-induced tilt and motion biases in relative horizontal velocity measured by a downward-looking ADCP mounted on a surface quasi-Lagrangian platform with a) hydrostatic and b) inertial responses. Analytical estimates with and without taking the ADCP beam response into consideration are shown in solid and dashed blue lines, respectively; the latter is the same as shown in Fig. 7 (blue line). Results of the semi-analytical simulation are shown for comparison (green dots).*

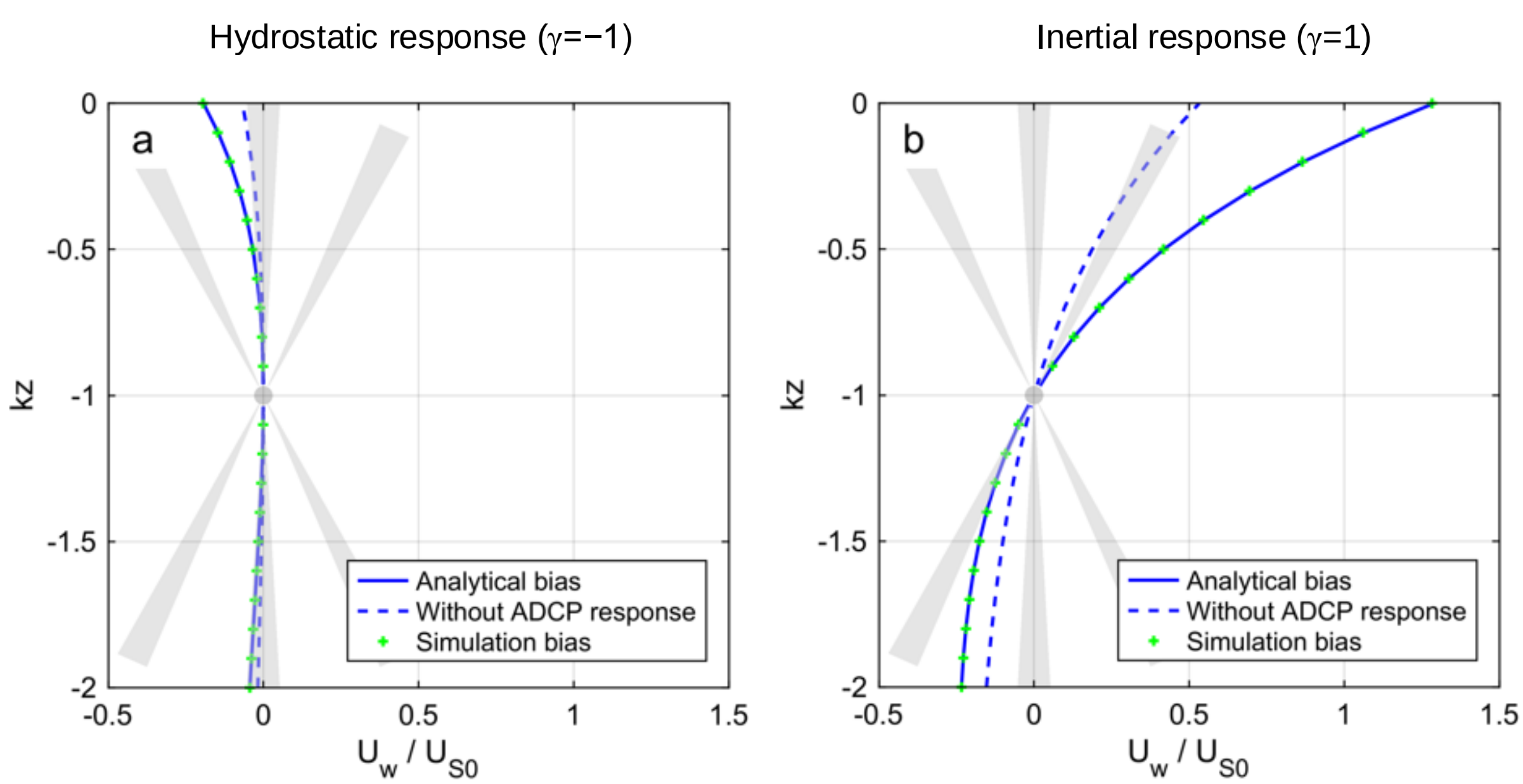

*Fig. 13 Same as Fig. 12, but for a subsurface quasi-Lagrangian instrument.*

## 5. Self-propelled platforms

The preceding sections focused exclusively on fully-Lagrangian platforms that follow the fluid particle motion. However, most real-life platforms are only partially Lagrangian in that they have a certain horizontal through-the-water speed. This propulsion can be intentional, as in the case of autonomous underwater or surface vehicles (AUVs and ASVs), as well as towed platforms, such as SWIMS (Klymak and Gregg 2001) or SurfOtter (Hughes et al. 2020). Unintentional propulsion can also arise from unavoidable windage of platforms that are meant to be Lagrangian, e.g., SWIFT (Thomson et al. 2015). Platform propulsion modifies the wave-induced biases in current observations, as shown by Amador et al. (2015)'s simulations.

Establishing the trajectory of a self-propelled platform affected by waves is not entirely straightforward, even for a simple case of a platform moving with a constant horizontal through-the-water velocity $u_p$. It is tempting to add $u_p$ to the horizontal component of the wave orbital velocity and integrate the trajectory in $(x, z)$ (e.g., as in Amador et al., 2017). However, this would result in the platform crossing the undulating isobaric (isopotential) surfaces. This is not realistic, as can be easily seen by considering a surface platform (see illustration in Fig. 14): with purely-horizontal propulsion, such platform would soon find itself in the air or underwater.

It is more realistic to assume that the platform propels itself along *isobaric* surfaces (which include the water surface). These surfaces are perpendicular to the effective gravity vector $\widehat{G}$ (see Fig. 5) and such propulsion can therefore be considered "locally-horizontal".
Physically, departure from an isobaric surface would be prevented by opposing pressure gradient forces (unless counteracted by the platform's control algorithm). We therefore include this "isobaric constraint" on the platform motion in subsequent calculations.

The self-propelled platform trajectory under isobaric constraint is described by a differential equation

$$\frac{dX_p}{dt} = u_p\widehat{H} - ia\omega e^{ikx_p - i\omega t + kz_0}, \tag{62}$$

where $\widehat{H} = i\widehat{G} = 1 + iake^{kz_0}\cos(kx_p - \omega t)$ is a unit vector in the local "horizontal" (isobaric) direction. This equation can be solved approximately by using the 0th order solution $x_p^{(0)} = x_0 + u_p t$ for $x_p$ on the right-hand side, producing

$$\frac{dX_p}{dt} = u_p + iaku_p e^{kz_0}\cos\phi_p - ia\omega e^{i\phi_p + kz_0}, \tag{63}$$

where $\phi_p = kx_0 - (\omega - ku_p)t$ is the 0th order approximation of the Doppler-shifted phase. This can now be integrated to

$$\begin{aligned} X_p &= X_0 + u_p t - ia\frac{ku_p}{\omega - ku_p} e^{kz_0} \sin\phi_p + a\frac{\omega}{\omega - ku_p} e^{i\phi_p + kz_0} = \\ &= X_0 + u_p t + a\frac{\omega}{\omega - ku_p} e^{kz_0} \cos\phi_p + iae^{kz_0} \sin\phi_p\,. \end{aligned} \tag{64}$$

Comparison of (64) and (3) shows that horizontal propulsion has a twofold effect on the wave-induced motion of the platform: Doppler-shifting of the wave frequency and scaling of the horizontal component of the wave-induced platform motion by a factor

$$\gamma_p = \frac{\omega}{\omega - ku_p} = \left(1 - \widetilde{u_p}\right)^{-1}, \tag{65}$$

where $\widetilde{u_p} = u_p/c_{ph}$ is the propulsion velocity normalized by the phase speed of the wave. For $0 < \widetilde{u_p} < 2$, horizontal excursions are enhanced; for $\widetilde{u_p} < 0$ or $\widetilde{u_p} > 2$ they are reduced. The value $u_p = c_{ph}$ is a singularity, corresponding to the platform "riding the wave." In this case, trajectories become aperiodic, and the effective velocity bias is equal to the orbital velocity $a\omega e^{kz_0}$, with an arbitrary direction depending on which part of the wave the platform is riding. However curious, this situation is not likely to occur in real life. When the propulsion speed is relatively slow ($\widetilde{u_p} \ll 1$), approximate relationship $\gamma_p \approx 1 + \widetilde{u_p}$ can be used.

The amplitude of the vertical excursion matches that of the water particles and remains constant regardless of the propulsion speed, which is a direct result of the isobaric constraint. Without such a constraint, both components of the wave-induced motions would be scaled by the same factor, as in the Amador et al. (2017) analysis. Such modification of the vertical excursion amplitude, however, is not realistic. This is, again, clearly demonstrated by a surface vehicle, whose vertical excursion cannot differ from that of a wave (see Fig. 14).

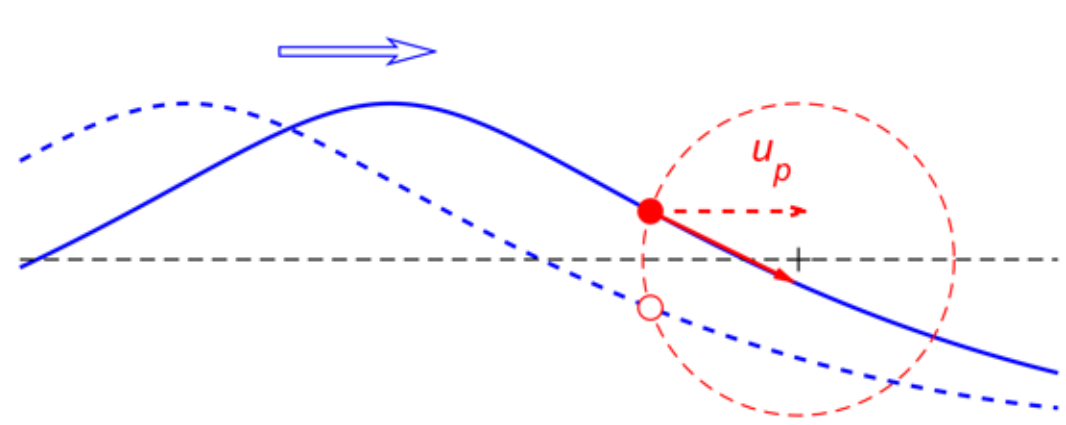

*Fig. 14 Illustration of isobaric constraint on the trajectory of self-propelled surface platform (solid red circle): In order for the platform to stay on the free surface (blue line), propulsion vector $u_p$ has to be parallel to the surface (solid red arrow) rather than purely horizontal (dashed arrow). A similar constraint applies to an underwater platform. Surface shape and the platform position at an earlier time is shown by dashed blue line and open red circle, respectively.*

Having established the platform trajectory, we can follow the same derivation steps as in Section 3.1. The expression for the absolute velocity in the sampling volume becomes

$$U_{wp} \approx U_1 + a^2\omega k e^{i\phi_p + k(z_1+z_0)}\left(\gamma_p \cos\phi_p - i\sin\phi_p\right) = U_1 + U_{wp} \tag{66}$$

Phase averaging leads to the expression for the wave-induced bias for a self-propelled platform:

$$\overline{U_{wp}} = \frac{1}{2}\left(\gamma_p + 1\right)a^2\omega k e^{k(z_1+z_0)}. \tag{67}$$

The factor $0.5\left(\gamma_p + 1\right)$ reflects the fact that the horizontal and vertical components of the elliptical sampling volume orbit contribute to the bias in proportion to their amplitudes, $\gamma_p a$ and $a$, respectively. This factor distinguishes our estimate of wave-induced bias from that obtained by Amador et al. (2017, eq. 10). As discussed earlier, in their analysis *both* components of the platform motion are scaled equally, which leads them to a factor of $\gamma_p$ in their eq. 10. Our application of the isobaric constraint on the platform trajectory avoids unrealistic scaling of the vertical amplitude, resulting in a factor of $0.5\left(\gamma_p + 1\right)$ instead. As expected, this factor equals 1 when if $u_p \ll c_{ph}$, and expression (67) reduces to (17).

As before, setting $z_1 = z_0$ gives the expression for Stokes drift experienced by a self-propelled platform:

$$U_{Sp} = \frac{1}{2}\left(\gamma_p + 1\right)a^2\omega k e^{2kz_0}. \tag{68}$$

Likewise, it can be shown that the same factor of $0.5\left(\gamma_p + 1\right)$ applies to other wave-induced biases arising in ADCP measurements conducted from a self-propelled platform.

It is important to note that the above analytical derivations use linear expansion for wave field velocity, which is only valid for relatively small platform excursions, $|X'| \ll k^{-1}$. As can be seen from (64), this assumption may be violated when $u_p$ approaches $c_{ph}$. For those cases, numerical integration of the trajectory equation (63) as implemented in the semi-analytical model, can be used. Fig. 15 shows the extent of the error of the analytical expression (68) compared to numerical integration. This error is less than 10%, unless $u_p$ is within 30% of $c_{ph}$ ($|1 - \widetilde{u_p}| < 0.3$, $|\gamma_p| > 3.3$).

Our analytical and numerical results differ from those obtained by Amador et al. (2015)'s simulations. For the same wave and platform speed parameters, eq. (68) predicts substantially lower values of $U_{Sp}/U_S$ than shown in their Fig. 2b (1.035 vs. ≈1.25). This discrepancy may be partially explained by Amador et al. using shallow water wave dispersion relationships. Additionally, it appears that their simulations do not use the isobaric constraint of the platform motion, as indicated by their reporting "magnification" of the vehicle's vertical motion for non-zero propulsion speeds. As mentioned earlier, this is only possible if the isobars are crossed. Amador et al. explain the origin of the bias as follows: "When the vehicle moves with the waves…, it prolongs its time in the crests leading to an alias of the average velocity in the direction of wave travel." This explanation is not supported by our analysis. As shown by eq. (64), wave-induced platform motion is vertically symmetric, i.e. the amount of time spent in the crests and in the troughs is the same In our analysis, the bias arises from the vertical modulation of orbital velocities, just like the regular (Lagrangian) Stokes drift.

In real-life applications, calculation of wave-induced bias should depend critically on the details of the response of the propulsion to the wave-induced motions – both in terms of the physical response (Does the thrust vector pitch with the pitching vehicle?), and the control system response (Does the vehicle attempt to maintain a level flight?). Therefore, applicability of our analytical expression (68) needs to be assessed on case-by-case basis.

An interesting case arises for platforms that are wave-driven, e.g., Boeing Liquid Robotics Wave Gliders (e.g., Mullison et al. 2011; Grare et al. 2021). The through-the-water propulsion of these platforms is unsteady, and it is likely to be coherent with the wave field to some extent. Although several dynamic models of Wave Gliders exist (e.g., Wang et al. 2019; Sun et al.

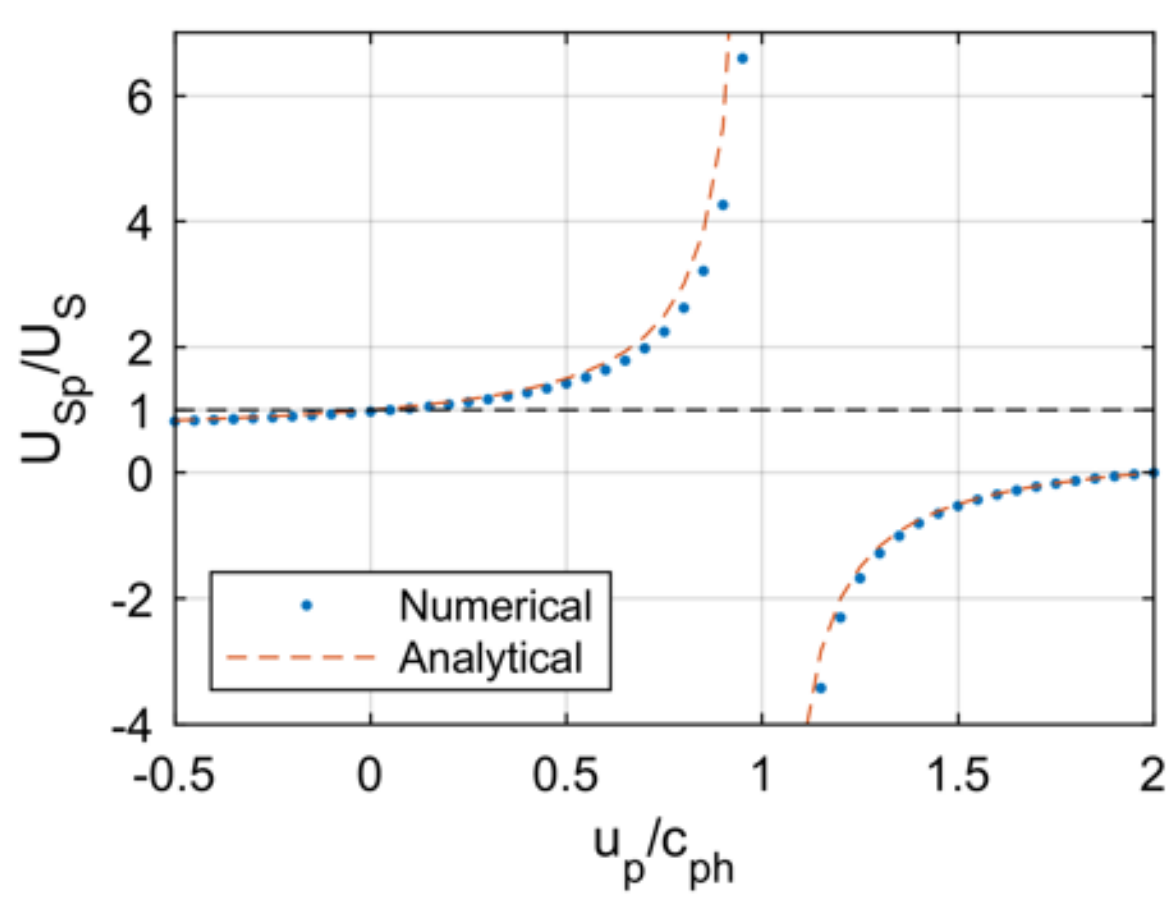

*Fig. 15 Stokes drift experienced by a self-propelled platform moving with the horizontal through-the-water speed $u_p$ based on numerical integration (dots) and the analytical expression (68).*

2022), we could not find a detailed discussion of the phase-resolved response of the vehicle to the wave field in the literature. Once the response is established, it should be possible to estimate any additional biases that may arise from the wave-coherent motion using numerical integration and the semi-analytical approach described in section 2.2.

## 6. Profiling platforms

Untethered profiling platforms, such as Lagrangian Float (MLF), are a promising solution for ocean velocity exploration due to their stability (Shcherbina et al. 2018, 2019). Closely related are the ADCP-equipped moored profilers (e.g., Zheng et al. 2022), although their motion is complicated by the interaction with the mooring line (see below). Underwater gliders (e.g., Ellis et al. 2015; Todd et al. 2017) also combine vertical profiling with horizontal propulsion discussed earlier.

Profiling platforms, like fully Lagrangian instruments, are subject to wave-induced biases. Similar to horizontally propelled platforms (section 5), corrections can be applied to account for vertical motion. These corrections, however, are not likely to be significant. Typical vertical speed of these profiling platforms, $w_p$, is relatively slow (from about 0.02 m s$^{-1}$ for MLF, to about 0.1 m s$^{-1}$ for underwater gliders, to 0.5 m s$^{-1}$ for moored profilers) in comparison with the typical phase speed of wind waves (5−30 m s$^{-1}$). The normalized profiling speed $\widetilde{w_p} = w_p/c_{ph}$ is therefore in the range of $10^{-3}$…$10^{-1}$. Just like in the case of horizontally propelled platforms, profiling corrections scale proportionately to $\widetilde{w_p}$, and are therefore small.

Profiling platforms are also subject to errors arising from averaging wave orbital velocities with increasing or decreasing amplitudes, which can be comparable to other wave-induced biases. These errors impose additional constraints on the velocity measurements from profiling platforms that will be discussed in greater detail in a subsequent publication.

## 7. Moored platforms

The motion of a moored current meters and ADCPs may be affected by the dynamics of the surface buoy as well as the mooring line. As originally discussed by Pollard (1973), vertical motion of the moored instruments is expected to be well coupled with that of the surface buoy; at the same time, horizontal motion is not well known, but likely more attenuated. Overall, eq. (17) and (52) can be viewed as upper bounds for the horizontal velocity bias induced by the motion of a moored current meter or an ADCP. It is important to note that $z_0 = 0$ now refers to the level of the surface buoy, as the buoy's motion dictates the motion of the instrument. If only the vertical motion of the buoy is transferred to the instrument, the bias is reduced by a factor of 2, as evident from (10). It is expected that the motion of the buoy should not affect the tilt of the moored instruments (unless they are mounted on the buoy itself), therefore tilt-induced biases should not be as important as in the case of untethered quasi-Lagrangian platforms.

Further research is needed to characterize the response of moored instruments to the wave field. With a better understanding of this response, the analytical and numerical techniques outlined here could be used to estimate the relevant wave-induced biases.

## 8. Finite-depth waves

All the dependencies derived in this paper use the deep-water (short-wave) approximation of the wave equations for simplicity. A more general expression for the elliptical finite-depth wave orbital trajectories is

$$\begin{aligned} x' &= a\frac{\cosh k(z_0+H)}{\sinh kH}\cos\phi_0 \\ z' &= a\frac{\sinh k(z_0+H)}{\sinh kH}\sin\phi_0 \end{aligned} \tag{69}$$

where $H$ is the water depth. Complex representation of these trajectories is less straightforward than (3), but can still be achieved by noting that an elliptical trajectory can be represented as a superposition of two counter-rotating circular orbitals, i.e.

$$a_1\cos\phi + ia_2\sin\phi = \frac{a_1+a_2}{2}e^{i\phi} + \frac{a_1-a_2}{2}e^{-i\phi}. \tag{70}$$

With this, we obtain

$$X' \equiv x' + iz' = \frac{a}{2\sinh kH}e^{i\phi_0+k(z_0+H)} + \frac{a}{2\sinh kH}e^{-i\phi_0-k(z_0+H)} = X'_+ + X'_-, \tag{71}$$

where

$$X'_\pm = \frac{a}{2\sinh kH}e^{\pm i\phi_0 \pm k(z_0+H)} \tag{72}$$

are two counter-rotating orbitals[5]. Physically, the two terms can be seen as two "virtual" deep-water waves that partially cancel each other to ensure no-normal-flow bottom boundary condition. In the deep-water limit $kh \gg 1$, expression (71) simplifies to (3).

[5] Note that (71) can also be written as $X' = a\sinh^{-1}(kH)\cosh(i\phi_0 + k(z_0+H))$, but it does not seem to offer additional insight nor simplify the calculations.

Using the generalized complex expression for the wave orbitals (67), the wave-induced biases can be calculated following the same steps as before. For example, expression for the motion-induced bias (17) becomes

$$U_w = \langle \omega k (X'_{1+} + X'_{1-})(X'^{*}_{0+} + X'^{*}_{0-}) \rangle. \tag{73}$$

The terms resulting from the products of counter-rotated orbitals, such as

$$X'_{1+} X'^{*}_{0-} = \frac{a^2}{4 \sinh^2 kH} e^{i\phi_0 + k(z_1+H)} e^{i\phi_0 - k(z_0+H)} X'_{1+} X'^{*}_{0-} = \frac{a^2}{4 \sinh^2 kH} e^{2i\phi_0 + k(z_1 - z_0)}, \tag{74}$$

are harmonic (with frequency $2\omega$) and vanish upon averaging over the wave phase. Remaining terms are non-periodic (just as in the deep-water case) and simplify to

$$\begin{aligned} U_w = \omega k (X'_{1+} X'^{*}_{0+} + X'_{1-} X'^{*}_{0-}) &= \frac{a^2 \omega k}{4 \sinh^2 kH} \left( e^{k(z_0+z_1+2H)} + e^{-k(z_0+z_1+2H)} \right) \\ &= a^2 \omega k \frac{\cosh k(z_0+z_1+2H)}{2 \sinh^2 kH}, \end{aligned} \tag{75}$$

as in Thomson et al. (2019).

Tilt-induced biases can be calculated similarly, with an important caveat: In finite-depth waves, effective anti-gravity vector is

$$-\hat{G} \approx i - ak \frac{\cosh k(z_0+H)}{\cosh kH} \cos \phi, \tag{76}$$

while the vertical material line vector is

$$\hat{M} \approx i + ak \frac{\sinh k(z_0+H)}{\sinh kH} \cos \phi_0. \tag{77}$$

(These formulae are derived similarly to (26) and (28) using the finite-depth dispersion relationship $\omega^2 = gk \tanh kH$.) While $-\hat{G}$ and $\hat{M}$ are still symmetric about the vertical near the surface ($z_0 = 0$), the symmetry breaks down towards the bottom ($z_0 = -H$), where the vertical material lines do not oscillate. Even though the platform tilt response now varies with depth, it can still be described by a universal formula similar to (29):

$$\hat{V} \approx i + \Upsilon(z_0) ak \cos \phi_0, \tag{78}$$

except that in this case the response function $\Upsilon(z_0)$ reflects the varying depth dependence of the platform response:

$$\Upsilon(z_0) = \begin{cases} -\dfrac{\cosh k(z_0+H)}{\cosh kH} & \text{(hydrostatic)} \\ +\dfrac{\sinh k(z_0+H)}{\sinh kH} & \text{(inertial)} \end{cases} \tag{79}$$

In the deep-water limit ($kH \gg 1$), $\Upsilon(z_0) = \gamma e^{kz_0}$ and (78) reduces to (29).

An expression for the tilt-induced "sweeping" bias then becomes

$$U_t = \langle \omega k X_1' r \hat{V}^* \rangle = \frac{1}{2} \Upsilon(z_0) a^2 \omega k^2 r \frac{\cosh k(z_1+H)}{\sinh kH}. \quad (80)$$

Similarly, the frame tilt bias is

$$U_f = \langle (U_1 - U_p)(i\hat{V}^*) \rangle = \frac{1}{2} \Upsilon(z_0) a^2 \omega k \frac{\sinh k(z_1+H) - \sinh k(z_0+H)}{\sinh kH}. \quad (81)$$

Note that the straightforward relationships among the various wave-induced biases that exist in the deep-water limit ((18), (34), (37)) do not hold in finite-depth case, except near the surface.

## 9. Broadband wave forcing

Even though the formulae derived so far are for a monochromatic wave, they can be readily generalized to account for a broadband wave field under the assumption of linear superposition. Given a directional spectrum of surface elevations $S_{\eta\eta}(\omega, \theta)$, as a function of cyclic frequency $\omega$ and propagation direction $\theta$ such that

$$\langle \eta^2 \rangle = \int_0^{2\pi} \int_0^{\infty} S_{\eta\eta}(\omega, \theta)\, d\omega\, d\theta, \quad (82)$$

the squared amplitudes of the spectral components can be expressed as

$$d(a^2) = 2 S_{\eta\eta}(\omega, \theta) d\omega\, d\theta. \quad (83)$$

Spectral expressions for various wave-induced biases are then obtained by integration of the respective monochromatic expressions. For example, the spectral expression for the absolute wave-induced bias due to platform motion becomes

$$\mathbf{U}_w \equiv (U_w, V_w) = 2 \int_0^{2\pi} \int_0^{\infty} \omega \mathbf{k} S_{\eta\eta}(\omega, \theta) e^{k(z_0+z_1)}\, d\omega\, d\theta, \quad (84)$$

where $\mathbf{k} \equiv (k_x, k_y) = k\hat{\mathbf{k}} = k(-\sin\theta, -\cos\theta)$ is the two-dimensional wavenumber[6]. Using the deep-water dispersion relationship $\omega^2 = gk$, this expression can be rewritten as

$$\mathbf{U}_w = 2 g^{-1} \int_0^{2\pi} \int_0^{\infty} \omega^3 \hat{\mathbf{k}} S_{\eta\eta}(\omega, \theta) e^{k(z_0+z_1)}\, d\omega\, d\theta. \quad (85)$$

Again, this expression can be related to the spectral representation of the Stokes drift profile (e.g., Liu et al. 2021):

$$\mathbf{U}_S = 2 g^{-1} \int_0^{2\pi} \int_0^{\infty} \omega^3 \hat{\mathbf{k}} S_{\eta\eta}(\omega, \theta) e^{2kz_0}\, d\omega\, d\theta. \quad (86)$$

Other bias expressions are obtained similarly.

[6] This expression assumes the wave propagation direction $\theta$ is specified as an "azimuth angle measured clockwise from true North to the direction wave is from", which is a standard for the National Data Buoy Center.

To illustrate the spectral behavior of the wave-induced bias, consider a broadband unidirectional (propagating in $x$-direction, $\theta$=270°) wave field with the frequency spectrum $S_{\eta\eta}(\omega)$. The discrete representation of such a wave field is

$$\eta = \sum a_n \sin(k_n x - \omega_n t), \tag{87}$$

with the amplitudes of individual components given by

$$a_n^2 = 2S_{\eta\eta}(\omega_n)\Delta_\omega, \tag{88}$$

where $\Delta_\omega$ is the spectral bandwidth. The corresponding equation for the bias due to the instrument motion is obtained by the summation of expression (17) for each component,

$$U_w = \sum a_n^2 \omega_n k_n e^{k_n(z_0+z_1)}. \tag{89}$$

Wind-generated wave spectra typically follow an $\omega^{-4}$ decay at frequencies higher than the wind-dependent peak, although the exact value of the exponent varies (Liu 1989). A realistic approximation of the wave field is therefore

$$a_n = a_0 \left(\frac{\omega_n}{\omega_0}\right)^{-2}, \tag{90}$$

where $\omega_0$ and $a_0$ are the peak wave frequency and amplitude, and $\omega_n > \omega_0$. The spectral behavior of the wave-induced bias due to the platform motion in this case is shown in Fig. 16. Similarly to the Stokes drift, the spectral components of the wave-induced bias decay more slowly with frequency ($\sim\omega_n^{-1}$) than the wave amplitude components (90). It is worth noting that although the "geometric mean" relationship between the profiles of the Stokes drift and the wave-induce bias (19) holds for the individual spectral components, it does not hold for their superposition, as can be seen by comparing (86) and (89).

Similarly, the broadband wave-induced bias due to platform tilt can be obtained by the summation of the monochromatic expression (34),

$$U_t = \sum \frac{1}{2}\gamma_n a_n^2 \omega_n k_n^2 r e^{k_n(z_0+z_1)}. \tag{91}$$

where the platform tilt response factor $\gamma_n$ is now a function of frequency. It is important to keep in mind that (34) was derived under the assumption of small platform pitch angle. Even if all the spectral components satisfy the small wave steepness assumption $a_n k_n \ll 1$, the net platform pitch amplitude may still exceed the small-angle approximation limits. If this is the case, bias estimates could be obtained by using a semi-analytical approach described in section 2.2, as it can handle arbitrarily high platform pitch angles.

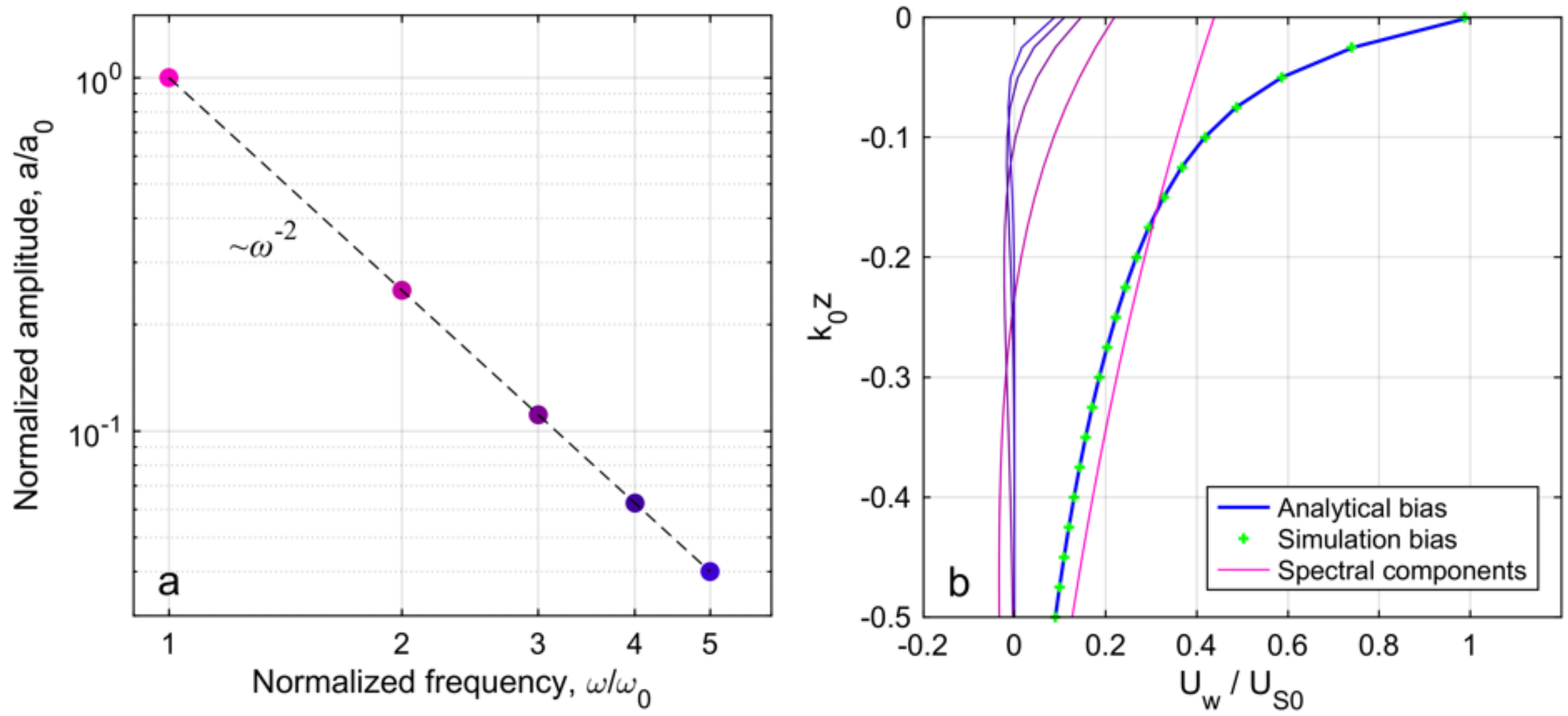

*Fig. 16 An illustration of the spectral behavior of the wave-induced bias in absolute horizontal velocity measured by a surface downward-looking Lagrangian instrument. (a) Normalized amplitudes of the 5 discrete wave modes approximating a wind-wave spectrum $S_{\eta\eta}$~$\omega^{-4}$. (b) Analytical estimates of the integral wave-induced bias (thick blue line) and its spectral components (color-coded by wave frequency). Results of the semi-analytical simulation are also shown for comparison (green dots). All velocity values are normalized by the integral surface Stokes drift; depth is normalized by the inverse lowest wavenumber, $k_0^{-1}$.Wave-induced instrument tilt is not included for simplicity. For reference, $U_{S0}$~35 cm s$^{-1}$ for fully-developed seas under 10 m s$^{-1}$ winds (Kenyon 1969).*

## 10.Prospects for mitigation

It is theoretically possible to mitigate some of the wave-induced biases with sufficient knowledge of the platform motion. Note, however, that the full mitigation requires the knowledge of *both* the wave field properties and the platform response characteristics, including the tilt response coefficient $\gamma$ (which is likely a function of frequency and direction of the wave forcing). Measuring directional wave spectra using these sensors is often a part of the autonomous platform's mission already (e.g., Thomson et al. 2018; Grare et al. 2021; Colosi et al. 2023), but additional efforts may be necessary to establish the platform response functions (D'Asaro and Shcherbina, in prep.)

Furthermore, results of our investigation suggest that it may be possible to correct the wave-induced biases by carefully measuring the motion of the platform, without the extraneous wave information. Autonomous sampling systems are routinely equipped with inertial attitude and heading reference systems (AHRSs), which could be further enhanced by GPS aiding for surface systems[7]. It is a widespread practice to apply instantaneous corrections for the platform motion and attitude to the ADCP observations on a single-ping basis before the necessary averaging is conducted (e.g., Grare et al. 2021). In itself, this approach compensates for the frame rotation bias, $U_f$, but not the motion or sweeping biases. However, the theoretical expressions derived above show common scaling of all the wave induced biases (Table 1). Therefore, functional relationships between $U_f$ and other wave-induced biases are expected to hold (e.g., see eq. (37)). There are caveats related to the tilt response coefficient, range dependence of the sweeping bias, as well as the ADCP response functions, which all enter these relationships. Nonetheless, it seems possible to determine $U_f$ as the difference between the velocity profiles measured with and without attitude corrections, and then use the theoretical considerations described in this paper to estimate other wave-induced biases. The feasibility of this approach is a subject of a follow-up study.

## 11. Conclusion

All the biases discussed above arise from the same basic mechanism – superposition of the wave orbital motions and the movement of the platform (and therefore the sampling volume). Since the two motions are at least partially coherent, their non-linear coupling produces aperiodic bias in the velocity measurements. These mechanisms can be seen as a generalization of the Stokes drift arising from the Lagrangian motion of a particle following wave orbital motions. In the case of Stokes drift, the “sampling volume” is the particle itself, and it moves in perfect synchrony with the “platform” (the particle). In a general case, the sampling volume is decoupled from the platform, as is the case for remote-sensing techniques (e.g., ADCP) or for a moored instrument pulled by the surface buoy (the case considered by Pollard, 1973). Therefore, in the case of monochromatic wave forcing, the biases scale as the geometric mean of the Stokes drift velocities at the nominal levels of the platform and the measurement, see eq. (19).

To first order, the biases arising from different components of the motion are independent and additive. The classic Stokes drift can be considered to be a sum of two identical biases, arising from the horizontal and vertical motions of the particle. In a general case considered here, there are two additional components arising from tilting of the platform. The sign of the latter, however, can differ depending on the details of the platform tilt response. All the biases discussed here impact only the horizontal component of the measured velocities, except in the case of a resonant (phase-shifted) platform response that may produce a vertical velocity bias (see section 3.3).

[7] It should be noted that we have considerable concerns regarding the dynamic accuracy of inexpensive low-power micro-electromechanical system (MEMS) AHRS implementations likely to be used on an autonomous platform, particularly with respect to resolving wave motions with periods of 2−10 s. This, however, is not an insurmountable issue, and it can be addressed by developing AHRS sensor-fusion algorithms optimized for capturing wave-induced motion. For more details, see D’Asaro and Shcherbina (in preparation).

ADCP beam geometry modifies the biases in a non-trivial way, introducing a number of velocity response functions (Section 4). We provide an analytical derivation of the response functions for a symmetric Janus beam configuration typical for 4- and 5-beam systems. Asymmetric and 3-beam systems would have different response functions, but they can be derived in a similar manner.

For a self-propelled platform navigating in the direction of wave propagation, the wave-induced biases grow dramatically as the platform's through-the-water speed approaches the phase speed of the wave. Conversely, a platform moving against the waves would experience a small attenuation of the wave-induced biases.

Analytical expressions for the three basic types of wave-induced biases in horizontal relative velocity sampling arising from quasi-Lagrangian platform motion, both with and without ADCP response functions taken into account, are shown in Table 1. In all cases, the wave-induced bias and the Stokes drift scale similarly with the wave amplitude and frequency but have different depth dependence. Unlike the Stokes drift, the relative wave-induced bias does not necessarily decay with depth. Overall, the surface Stokes drift can be used as an order-of-magnitude upper-limit estimate of the wave-induced bias. For fully-developed seas under 10 m $s^{-1}$ winds, this value would be ~35 cm $s^{-1}$ (Kenyon 1969).

In principle, it is possible to compute all the biases affecting an observational dataset as long as the properties of the wave field and the platform response are known. In practice, however, these parameters may not be known with sufficient detail. Nonetheless, the analytical expressions derived here (Table 1) can be used for estimating the observational uncertainties arising from wave-induced effects. These formulae may also be useful during the observing system design and experiment planning phases, as they show the tradeoffs associated with various instrument and platform configurations. Broadly speaking, an upward-looking ADCP mounted on a subsurface quasi-Lagrangian platform with hydrostatic response can be expected to have weaker wave-induced biases when observing velocities at a given depth than other configurations (Fig. 13a). Naturally, these recommendations are not a hard and fast rule, and they need to be considered in context of broader experimental requirements.

The results of this study are expected to be generalizable beyond the linear surface waves. Stokes drift is known to occur in a wide variety of non-transverse waves (Guha and Gupta 2024), of which internal gravity waves may be particularly interesting (e.g., Coy et al. 1986). It is therefore reasonable to anticipate that phenomena similar to wave-induced biases described here may also be present in other contexts. From the illustration in Fig. 1, it could be summarized that the necessary "ingredients" of the wave-induced bias are 1) variation of velocities (or other variable being measured) in the wave phase space, and 2) particular coherent motion of the sampling volume through this space. (Note that while the evanescent vertical dependence of the wave amplitude contributed to the biases discussed here, it is not strictly necessary for these biases to occur.) These two necessary factors can be found in a broad range of situations, sometimes seemingly unrelated to velocity measurements. E.g., Franks et al. (2020) demonstrated that the onshore transport of plankton by internal gravity waves depends on whether the plankton is neutrally buoyant or depth-keeping. If we consider plankton as a 'velocity sampler' with a particular response to wave-induced motions, the conceptual parallels between Franks et al. (2020) and our study become evident. Therefore, we hope that the analytical and numerical frameworks developed here will provide broader insights into wave-induced effects across various contexts.

| Wave-induced bias | "Vector" sampler | ADCP |
|---|---|---|
| Motion ($U_{wr}$) | $U_{S0}e^{kz_0}(e^{kz_1} - e^{kz_0})$ | $U_{S0}e^{kz_0}(R_u e^{kz_1} - e^{kz_0})$ |
| Sweeping ($U_t$) | $\frac{1}{2}\gamma U_{S0}kr\, e^{k(z_1+z_0)}$ | $\frac{1}{2}\gamma U_{S0}kr\, e^{k(z_1+z_0)}R_t$ |
| Frame rotation ($U_f$) | $\frac{1}{2}\gamma U_{S0}e^{kz_0}(e^{kz_1} - e^{kz_0})$ | $\frac{1}{2}\gamma U_{S0}e^{kz_0}(R_u e^{kz_1} - e^{kz_0})$ |

*Table 1 Analytical expressions for various wave-induced biases in horizontal relative velocity sampling with a "Vector" sampler (disregarding the ADCP response), and an ADCP. All the biases scale with the surface Stokes drift, $U_{S0} = a^2\omega k$. Tilt response factor $\gamma = -1$ corresponds to hydrostatic platform response mode, $\gamma = 1$ corresponds to inertial response (Section 3.3). Response functions $R_u$ and $R_t$ are determined by the ADCP beam geometry (Section 4). Modification of the formulas for self-propelled platforms and finite-depth wave equations is discussed in Sections 5 and 8, respectively. See text for other definitions.*

***Acknowledgments:*** We thank J. Tom Farrar and Jim Thomson for many fruitful discussions of this topic over the years. We also thank Dr. Gilberto Javier Fochesatto, Dr. Nick Pizzo and an anonymous reviewer for their thorough reading of the manuscript and numerous useful comments and suggestions. This work was supported by the Office of Naval Research "CALYPSO" and "Langmuir" Departmental Research Initiatives (N00014-18-1-2139, N00014-18-1-2420, N00014-14-1-0666, N00014-17-1-2859), and by NASA S-MODE project, an EVS-3 Investigation awarded under NASA Research Announcement NNH17ZDA001N-EVS3.

***Availability* Statement:** No datasets were generated or analyzed during the current study**.** MATLAB implementation of the semi-analytical model is freely available at https://github.com/shcher2018/wave-bias under MIT license.